\documentclass[aps,twocolumn,superscriptaddress,amsmath,amssymb]{revtex4-1}

\usepackage{graphicx}
\usepackage{dcolumn}
\usepackage{bm}

\usepackage{subcaption}
\usepackage{color}

\newcommand{\beq}{\begin{equation}}
\newcommand{\eeq}{\end{equation}}
\newcommand{\ba}{\begin{array}{ccc}}
\newcommand{\ea}{\end{array}}
\newcommand{\nn}{\nonumber \\}

\def\bea{\begin{eqnarray}}
\def\eea{\end{eqnarray}}

\begin{document}

\preprint{APS/Ising-KT}

\title{Emergent supersymmetry at the Ising-Berezinskii-Kosterlitz-Thouless multicritical point}

\author{Liza Huijse}
 \email{lhuijse@stanford.edu}
\affiliation{Stanford Institute for Theoretical Physics, Stanford University, Stanford, California 94305, USA}
\affiliation{Department of Physics, Harvard University, Cambridge, Massachussetts 02138, USA}

\author{Bela Bauer}
\affiliation{Station Q, Microsoft Research, Santa Barbara, California 93106, USA}

\author{Erez Berg}
\affiliation{Department of Condensed Matter Physics, Weizmann Institute of Science, Rehovot, 76100, Israel}
\affiliation{Department of Physics, Harvard University, Cambridge, Massachussetts 02138, USA}

\date{\today}

\begin{abstract}
We show that supersymmetry emerges in a large class of models in $1+1$ dimensions with both $\mathbb{Z}_2$ and $U(1)$ symmetry at the multicritical point where the Ising and Berezinskii-Kosterlitz-Thouless transitions coincide. To arrive at this result we perform a detailed renormalization group analysis of the multicritical theory including all perturbations allowed by symmetry. This analysis reveals an intricate flow with a marginally irrelevant direction that preserves part of the supersymmetry of the fixed point. The slow flow along this special line has significant consequences on the physics of the multicritical point. In particular, we show that the scaling of the $U(1)$ gap away from the multicritical point is different from the usual Berezinskii-Kosterlitz-Thouless exponential gap scaling.
\end{abstract}


\maketitle

\begin{figure}
\includegraphics[width=\columnwidth]{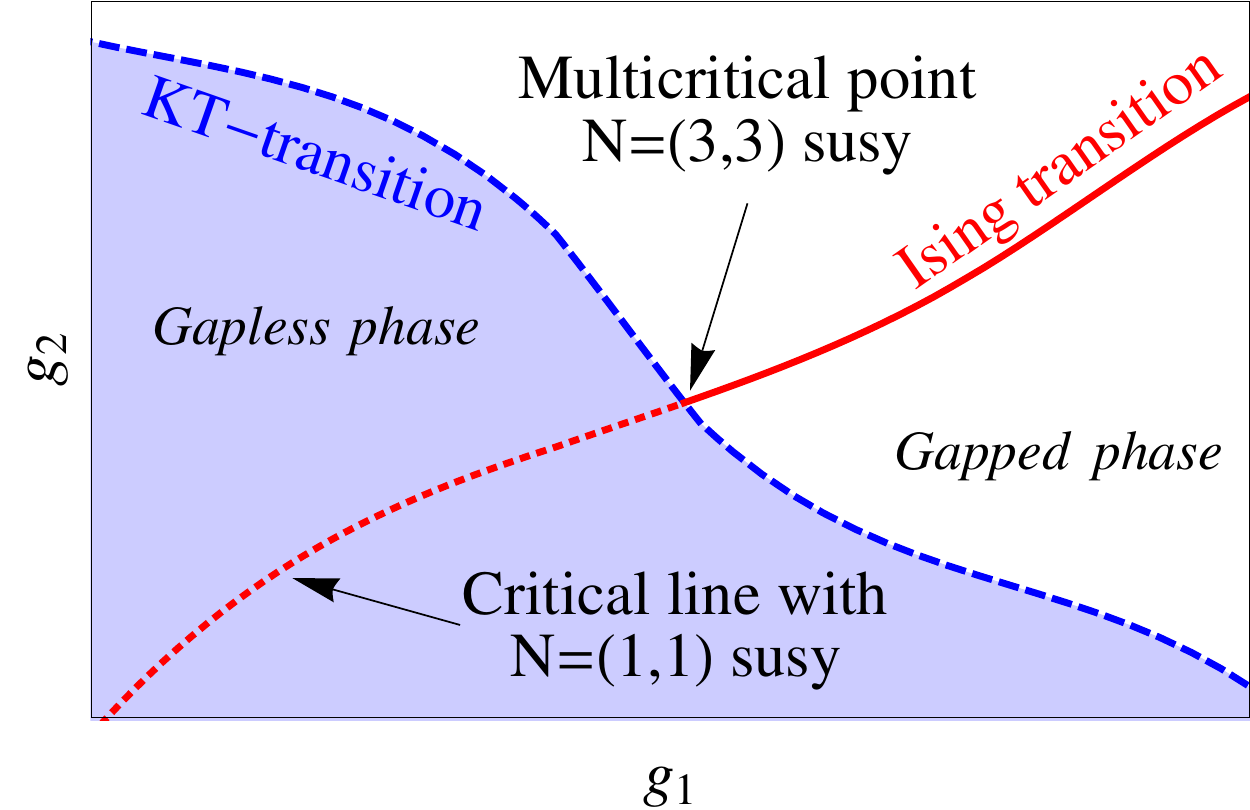}
\caption{Schematic phase diagram for two parameters, $g_1$ and $g_2$. The blue dashed line is the Kosterlitz-Thouless transition, corresponding to $K=K_{\textrm{crit}}$, separating the gapless quasi long-ranged order phase (shaded region, $K>K_{\textrm{crit}}$) from the gapped disordered phase. The red line is the Ising transition. Inside the gapless phase the Ising transition can have $\mathcal{N}=(1,1)$ supersymmetry (dotted red line). The multicritical point is located at the intersection of the BKT and Ising transitions. \label{fig:PD}}
\end{figure}

\textit{Introduction.--}Characterizing and classifying the transitions between distinct phases of matter
has been a perennial question in physics, but also one of its greatest successes
due to the emergence of universal behavior at continuous phase transitions. At
such transitions, the low energy physics is independent of microscopic details and can be fully understood in terms of its {\it universality class}. Famous examples of universality classes in two dimensional systems are the Ising transition associated to the breaking of a $\mathbb{Z}_2$ symmetry and the Berezinskii-Kosterlitz-Thouless (BKT) transition in systems with a $U(1)$ symmetry. In this paper we study the properties of the multicritical point where an Ising and BKT transition coincide in one-dimensional quantum or two-dimensional classical systems (see Fig. \ref{fig:PD}). We find that under certain conditions, a new universality class emerges at this multicritical point, which is characterized by emergent supersymmetry and qualitatively new physical behavior.

Supersymmetry pertains to the invariance of the system under a transformation that maps bosons into fermions and vice versa. It plays an important role in high energy physics, where it may provide a solution to the hierarchy problem. However, supersymmetry can also play a role in condensed matter systems, as an explicit or an emergent symmetry \cite{Friedan84,Foda,fendley2003,fendley2003-1,Veneziano06c,huijse2008,SSLee,Grover12,hagendorf2012}. To the best of our knowledge, the Ising-BKT multicritical point is the first example of a condensed matter system where extended ($\mathcal{N}>1$) supersymmetry emerges. 

This multicritical point can, in principle, occur in any system with both a $\mathbb{Z}_2$ and a $U(1)$ symmetry.
Possible realizations include one-dimensional spin chains where the $SU(2)$ spin symmetry is broken down to a $U(1)\otimes \mathbb{Z}_2$ symmetry, and polar molecules \cite{Jin13} in an optical lattice confined to one dimension by a non-circularly symmetric potential. In the latter example, if the dipoles are oriented by an external field, the strong repulsive interactions between the polar molecules give rise to a ``zig-zag'' instability, where the molecules move slightly away from the potential minimum at the center of the confining potential to form an alternating pattern that increases their distance \cite{Kollath08, Ruhman11, Abhijit13, Gammelmark13}. In the zig-zag phase, a $\mathbb{Z}_2$ symmetry associated with reflection is spontaneously broken. In addition, if the density of the molecules is commensurate with respect to the optical lattice along the system, a BKT superfluid to Mott insulator occurs at a critical strength of the lattice. The multicritical point occurs when the lattice strength and the transverse confining potential are tuned such that these two transitions coincide.

We show that in a large class of systems this multicritical point is described by a Lorentz-invariant theory with $\mathcal{N}=(3,3)$ supersymmetry, where both the Lorentz-invariance and the supersymmetry are emergent properties. We arrive at this result by a detailed renormalization group (RG) analysis of the decoupled multicritical theory and its symmetry-allowed perturbations. Interestingly, we find a very slow flow towards the decoupled Ising-BKT theory, as a result of marginally irrelevant operators. These operators give rise to qualitatively new behavior, such as an unusual scaling of the correlation length away from the multicritical point.

Already in the late 80s, Foda \cite{Foda} considered two-dimensional classical systems with a $U(1)$ and $\mathbb{Z}_2$ symmetry and found emergent $\mathcal{N}=(1,1)$ supersymmetry at the Ising transition provided it occurs at a lower temperature than the KT transition. The absence of a coupling between the two theories and emergence of Lorentz-invariance is ensured by a $90^o$ rotational symmetry of the systems that are considered. Recently, Sitte et al. \cite{Sitte} revealed an intricate RG flow diagram for this theory when the constituent theories are coupled. We point out that their result in combination with the work of Foda implies that even systems lacking the $90^o$ rotational symmetry may exhibit a critical point with emergent $\mathcal{N}=(1,1)$ supersymmetry (see fig. \ref{fig:PD}). Note that in a two dimensional phase diagram this corresponds to a critical line that lies within the gapless phase of the $U(1)$ sector. The present work is a natural extension of these results to the multicritical point where supersymmetry is enhanced.

\textit{Model.--}We consider the Lagrangian density
$\mathcal{L}_0=\mathcal{L}_{\Phi}+\mathcal{L}_{\chi}$,
where
\begin{eqnarray}
\mathcal{L}_{\Phi}&=&\frac{1}{2 \pi K} \Big( \frac{1}{v}(\partial_{\tau}\Phi)^2 + v(\partial_{x}\Phi)^2 \Big) \nonumber\\
\mathcal{L}_{\chi}&=&\chi_{R}(\partial_{\tau}+\frac{u}{\imath}\partial_{x})\chi_{R}+\chi_{L}(\partial_{\tau}-\frac{u}{\imath}\partial_{x})\chi_{L}. \label{eqn:L0}
\end{eqnarray}
Here, $v,K$ are the velocity and Luttinger parameter of the
bosonic field, $\Phi$, and $u$ is the velocity of the left- and right-moving modes of the real (Majorana) fermionic field, $\chi$. If Lorentz invariance is present, the fermion and boson velocities are equal, $u=v$. In this case, the resulting conformal field theory has an $\mathcal{N}=(1,1)$ supersymmetry generated by the supercharges $\chi_L \partial_z \Phi_L$ and $\chi_R \partial_{\bar{z}} \Phi_R$, where $z=x + \imath v \tau$ and $\Phi(z,\bar{z})=\Phi_L(z)+\Phi_R(\bar{z})$. For the special case of $K=4$, there is an additional extended supersymmetry, an $\mathcal{N}=(2,2)$ supersymmetry, generated by the supercharges $\chi_L \exp(\pm \imath \sqrt{2} \Phi_L)$ for the left-movers and similarly for the right-movers (more details are provided in Appendix \ref{sec:susy}).

Adding all perturbations to the fixed point action that are allowed if only $\mathbb{Z}_2$ and $U(1)$ symmetry are imposed, leads to
\begin{eqnarray}
\mathcal{L}_{\rm{int}}&=& \imath m \chi_R \chi_L  + g \cos(\sqrt{2} \Phi) -\lambda (\partial_x \Phi) \imath \chi_R \chi_L. \label{eqn:L}
\end{eqnarray}
The first term, the mass term for the fermion, is a relevant term that takes us to the Ising ordered or disordered phase. At the multicritical point, we have $m=0$. The cosine-term for the boson has scaling dimension $h=K/2$. At the multicritical point, that is for $K=K_{\textrm{crit}}=4$, this term is marginal. The last term is a marginal term that couples the boson to the fermion \cite{Sitte}. Note that this term explicitly breaks Lorentz symmetry. However, in 1+1 dimensional lattice systems (such as the ones described above), this symmetry is typically absent, and this term is allowed. All other perturbation are either irrelevant or not allowed by symmetry. Finally, the fermion and boson velocities in the fixed point action are allowed to be different.

\textit{RG analysis.--}The renormalization group flow for $\mathcal{L}=\mathcal{L}_{\Phi}+\mathcal{L}_{\chi}+\mathcal{L}_{\rm{int}}$ for $g=0$ was worked out in Ref.~\cite{Sitte}. In \cite{Bauer13} we derived the flow equations to second order in the couplings, in the presence of the cosine term. In a scheme where the boson velocity is kept fixed and an anomalous dynamical exponent, $z$, is introduced, it was found that
\begin{eqnarray}
\frac{du}{d\ell}&=&-\frac{u\lambda^{2}K}{4}\Big(\frac{1}{(v+u)^{2}}-\frac{1}{4uv}\Big)\nonumber\\
\frac{dm}{d\ell}&=& m \left( 1- \frac{\lambda^2 K}{8} \left[ \frac{1}{u(u+v)}+\frac{1}{(u+v)^2}-\frac{1}{2uv} \right] \right) \nn
\frac{d\lambda}{d\ell}&=&0\nonumber\\
\frac{dK}{d\ell}&=&-K^{2}\Big(\frac{g^{2}\pi^2}{2v^{2}}-\frac{\lambda^{2}}{16 uv}\Big)\nonumber\\
\frac{dg}{d\ell}&=&\frac{g}{2}(4-K)\nonumber
\end{eqnarray}
and
\bea
z&=&1+\frac{\lambda^{2}K}{16 uv}.\nonumber
\eea
For $\lambda=0$, the equations for $g$ and $K$ reduce to the Kosterlitz equations. There is a line of fixed points to second order in the couplings parametrized by
\begin{eqnarray}
\lambda=2\sqrt{2}\pi g, \ K=4, \ m=0, \ \textrm{and}\ u=v. \nonumber
\end{eqnarray}
Remarkably, as shown in \cite{Bauer13}, precisely this line preserves part of the $\mathcal{N}=(2,2)$ supersymmetry. For a small velocity difference we find, writing $u-v=\epsilon$,
\begin{eqnarray}
\frac{d\epsilon}{d\ell}&=&\frac{\lambda^{2}K}{64 v^3}\epsilon^2,\nonumber
\end{eqnarray}
such that the velocity difference is relevant for $u>v$, and irrelevant for $u<v$ (this is also true for $g=0$ \cite{Sitte}). To determine whether for $u \leq v$ the flow at the multicritical point is towards the fixed point, i.e. $\lambda=0$, we are faced with the rather challenging task of computing the flow equation for $\lambda$ to third order. 

We proceed by fermionizing the boson and then, taking a field theoretic approach, we introduce counter terms for the bare couplings and compute the Callan-Symanzik equation at the cutoff scale. To fermionize the boson we introduce a second, redundant bosonic field $\Phi_s$, with velocity $v_s$ and Luttinger parameter $K_s$, which is free and completely decoupled from the other fields. The theory is fermionized by introducing two flavours of complex fermions, $\psi_{a,p}$ where $a=1,2$ and $p=\pm$ for right and left movers, respectively. The precise relation between $\psi_{a,p}$ and the bosonic fields $\Phi$, $\Phi_s$ is given in Appendix \ref{sec:fermionize}. We then find, by reverse engineering the usual bosonization steps, that the fermionized action reads $\mathcal{L}=\mathcal{L_{\psi}}+\mathcal{L_{\chi}}+\mathcal{L_{\rm{int}}}$, with $\mathcal{L_{\chi}}$ as defined above,
\bea
\mathcal{L_{\psi}}&=&\sum_{p=\pm} \sum_{a=1,2}\psi_{a,p}^{\dagger}(\partial_{\tau}- \imath p v_F \partial_{x})\psi_{a,p}, \nonumber
\eea
and
\begin{widetext}
\begin{eqnarray}
\mathcal{L_{\rm{int}}}=  \imath m \chi_R \chi_L  + 2 \pi^2 g \left( \psi_{1+}^{\dagger} \psi_{1-}  \psi_{2+}^{\dagger} \psi_{2-} +\psi_{1-}^{\dagger} \psi_{1+}  \psi_{2-}^{\dagger} \psi_{2+} \right) +\bar{g}  \sum_{a=1,2} \left( \psi_{a,+}^{\dagger}\psi_{a,+} \right)  \sum_{a=1,2}  \left( \psi_{a,-}^{\dagger}\psi_{a,-} \right)  \nn
 - \sqrt{2} \pi \lambda   \sum_{\substack{p=\pm \\ a=1,2}} \left( \psi_{a,p}^{\dagger}\psi_{a,p} \right) \imath \chi_R \chi_L  +2 g_v \sum_{p=\pm} \psi_{1,p}^{\dagger}\psi_{1,p} \psi_{2,p}^{\dagger}\psi_{2,p} - \sum_{\substack{p=\pm \\ a=1,2}} \psi_{a,p}^{\dagger}( \imath p \epsilon_v \partial_{x})\psi_{a,p} + \imath \epsilon_u(\chi_{L} \partial_{x}\chi_{L}- \chi_{R} \partial_{x}\chi_{R}) ,\nonumber  \label{eqn:Lferm}
\end{eqnarray}
\end{widetext}
where the last three terms are a forward scattering term and velocity renormalizations for the fermion velocities $v_F$ and $u$. These terms are included because they are generated by the interaction terms under RG. The parameters of the bosons are related to those of the complex fermions as follows: $v=v_F+\epsilon_v+g_v/\pi$, $K=4(1-\bar{g}/(\pi v_F))$, $v_s=v_F+\epsilon_v-g_v/\pi$, and $K_s=1$.


We refer to appendix \ref{sec:2ndorder} for the computation and verification of the RG equations to second order in the fermionized model. We proceed to compute the beta function for $\lambda$ to third order in the couplings on the special line that does not flow to second order. We will argue below that this suffices to establish stability of the fixed point. Schematically, we first perturbatively compute the $4$-point function, $G_{\lambda}^{(4)} =\langle \psi_{1+}^{\dagger}\psi_{1+} \chi_R \chi_L \rangle$, to third order in the couplings. We then introduce counter terms for the divergent diagrams by imposing the renormalization condition at the renormalization scale $M$. Finally, we use the fact that the $4$-point function obeys the Callan-Symanzik (CS) equation to obtain the beta function:
\bea
\left[ M \frac{\partial}{\partial M} - \sum_i \beta(g_i) \frac{\partial}{\partial g_i} +2 \gamma_{\psi} + 2 \gamma_{\chi} \right] G_{\lambda}^{(4)}=0, \nonumber
\eea
where $g_i$ are all the couplings, $\beta (g_i )=-\partial g_i/\partial (\ln M)$ and $\gamma_{\psi,\chi}$ are the field renormalizations. Note that we have defined the beta functions with the sign convention commonly used in the condensed matter community (see, e.g., \cite{Shankar}), which is the opposite sign convention to that typically used in field theory textbooks (e.g. 
\cite{Peskin}).

\begin{figure}
\centering

\begin{subfigure}{.5\columnwidth}
  \centering
  \includegraphics[width=.9\columnwidth]{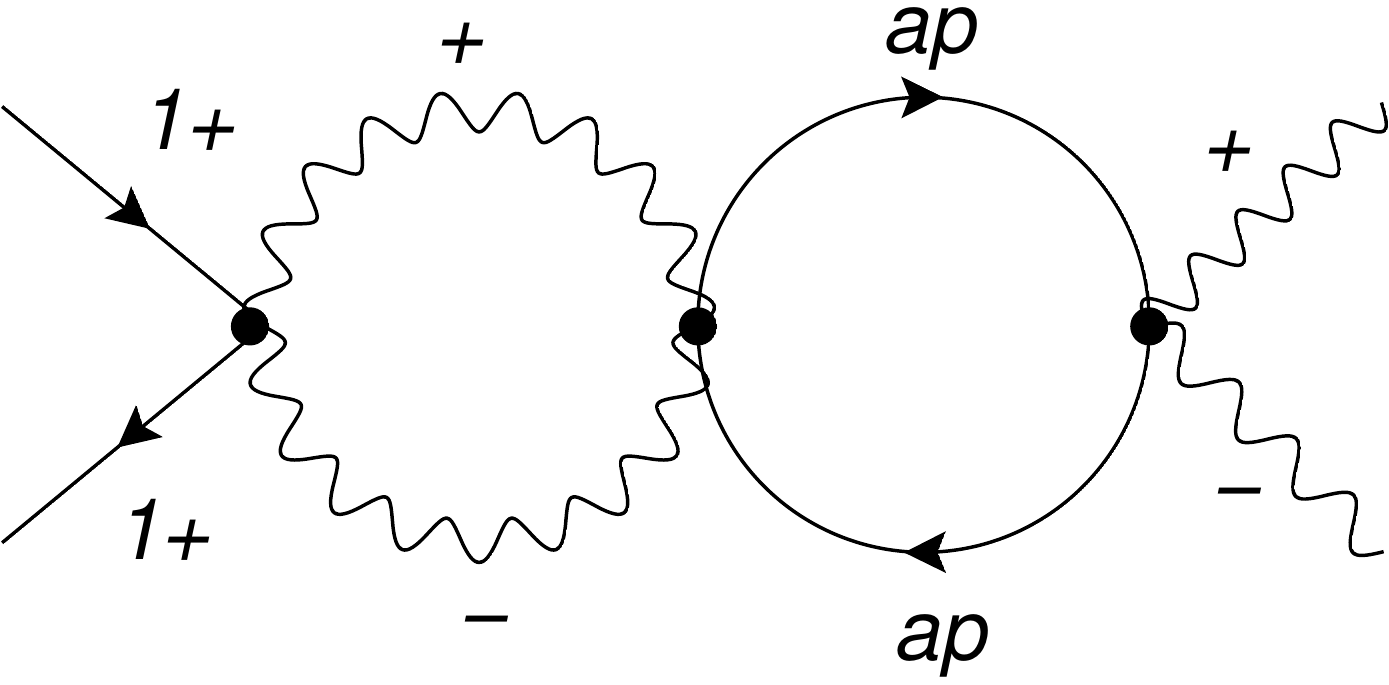}
  \caption{$\lambda^3$ \label{sf:d2}}
\end{subfigure}%
\begin{subfigure}{.5\columnwidth}
  \centering
  \includegraphics[width=.9\columnwidth]{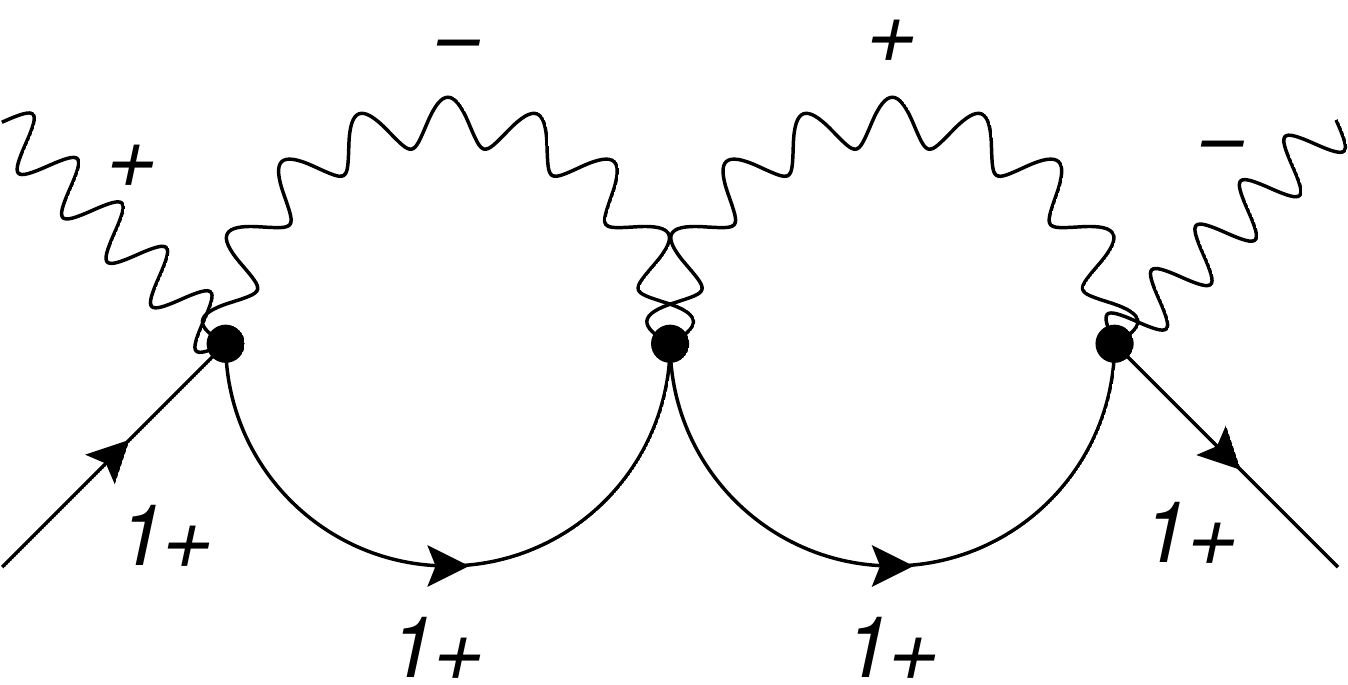}
  \caption{$\lambda^3$ \label{sf:d6}}
\end{subfigure}\\

\begin{subfigure}{.3\columnwidth}
  \centering
  \includegraphics[width=.9\columnwidth]{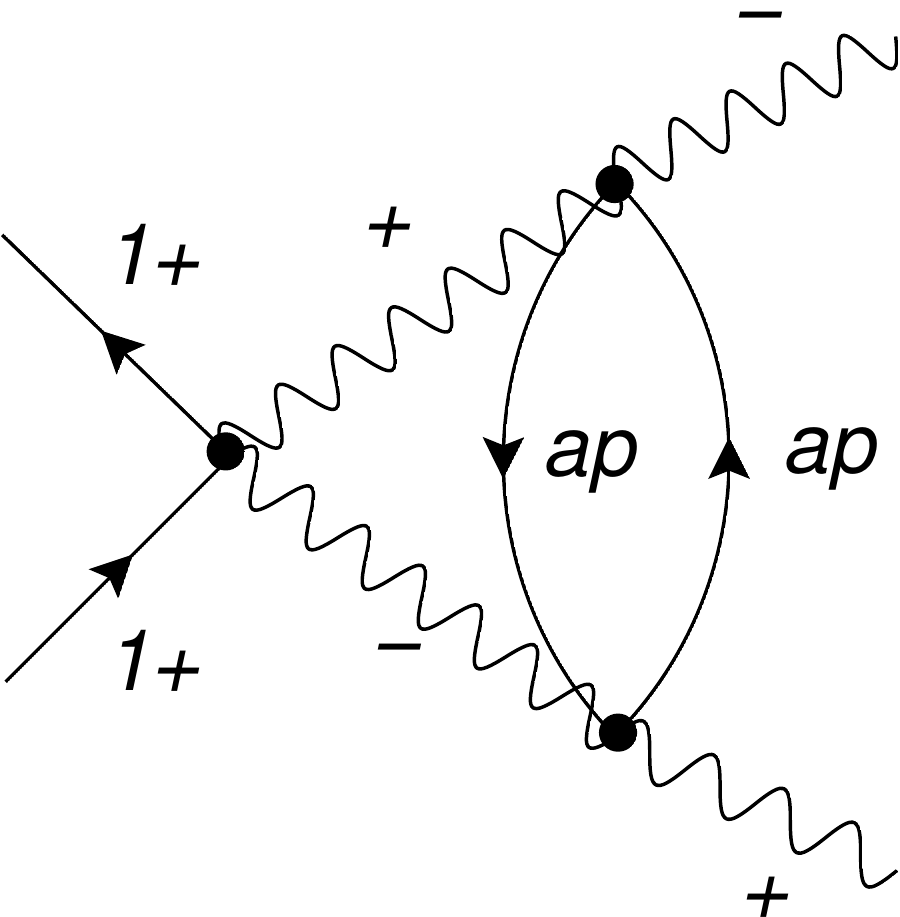}
  \caption{$\lambda^3$ \label{sf:d1}}
\end{subfigure}%
\begin{subfigure}{.3\columnwidth}
  \centering
  \includegraphics[width=.9\columnwidth]{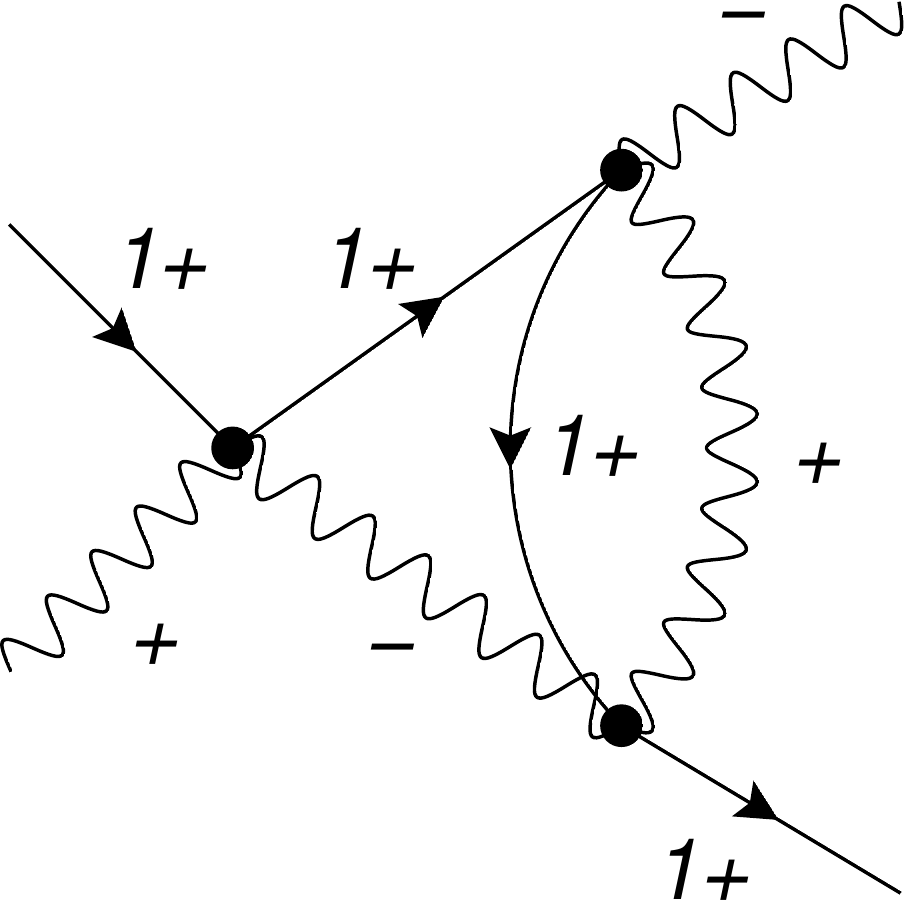}
  \caption{$\lambda^3$  \label{sf:d7}}
\end{subfigure}
\begin{subfigure}{.3\columnwidth}
  \centering
  \includegraphics[width=.9\columnwidth]{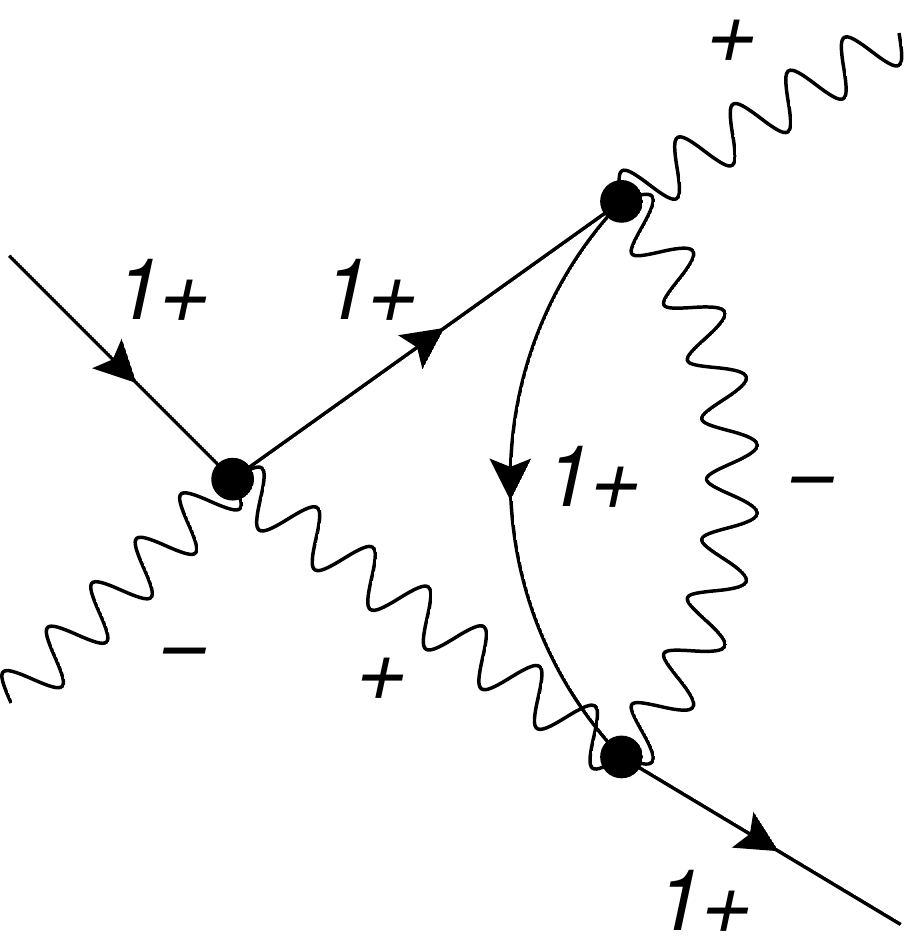}
  \caption{$\lambda^3$  \label{sf:d8}}
\end{subfigure}\\

\begin{subfigure}{.3\columnwidth}
  \centering
  \includegraphics[width=.9\columnwidth]{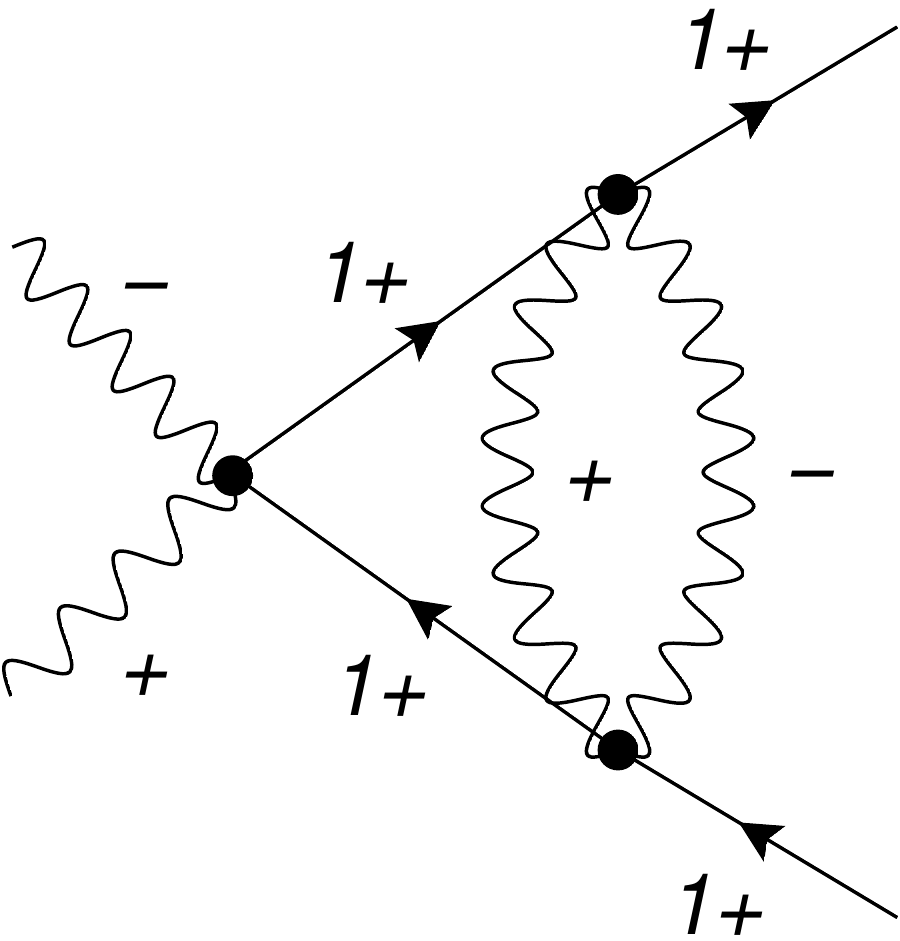}
  \caption{$\lambda^3$  \label{sf:d3}}
\end{subfigure}
\begin{subfigure}{.3\columnwidth}
  \centering
  \includegraphics[width=.9\columnwidth]{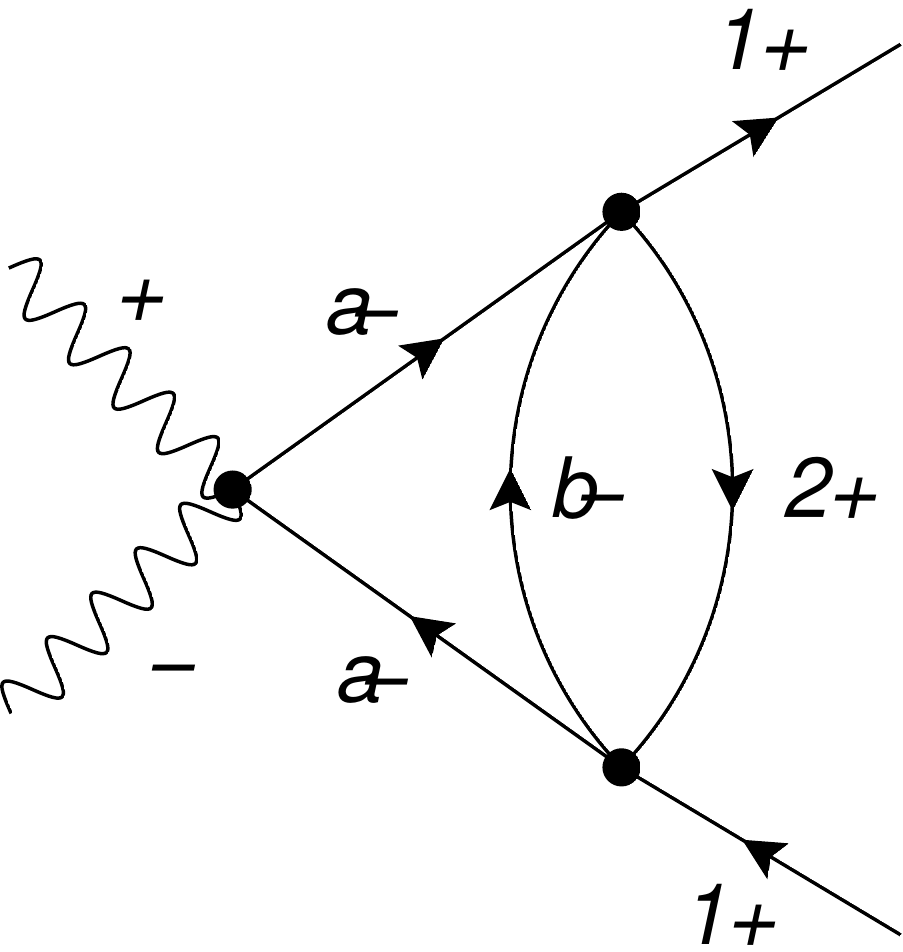}
  \caption{$\lambda g^2$  \label{sf:d4}}
\end{subfigure}
\begin{subfigure}{.3\columnwidth}
  \centering
  \includegraphics[width=.9\columnwidth]{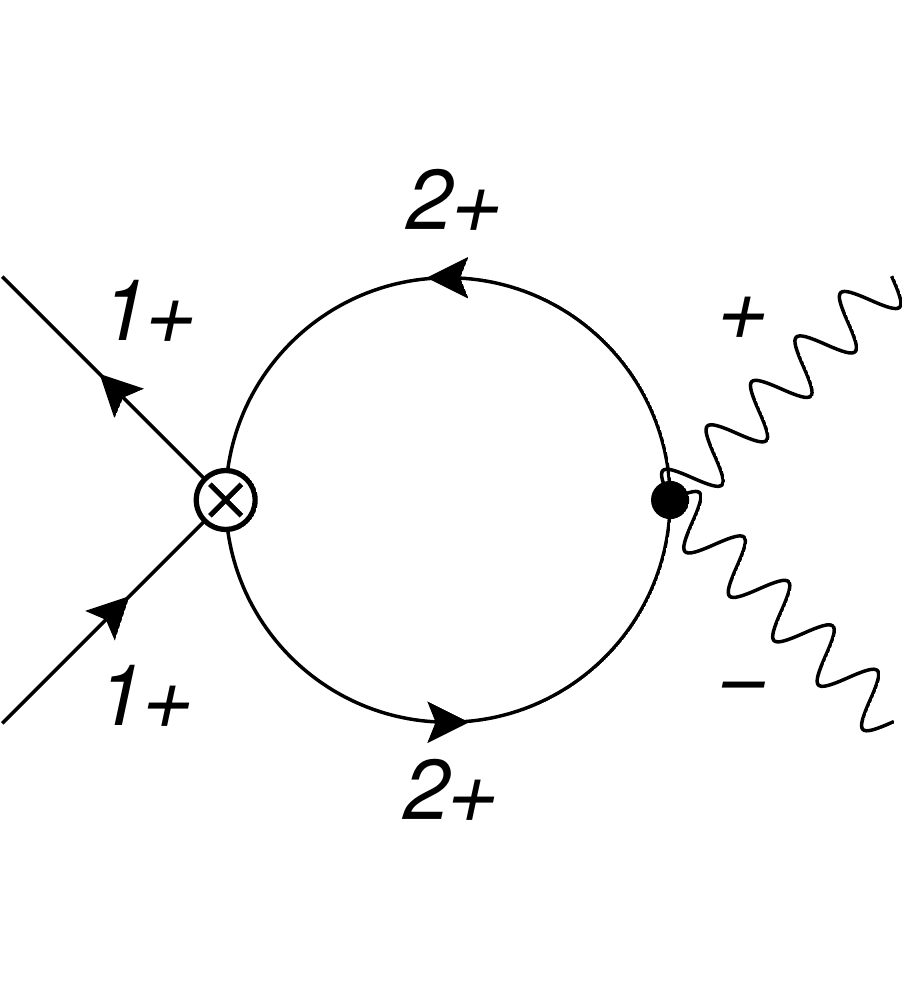}
  \caption{$\delta_{g_v} \lambda$ \label{sf:lgv}}
\end{subfigure}

\caption{The divergent diagrams that contribute to the $\lambda$-counter term at third order, where $a=1,2$, $p=\pm$ and $b=1,2$ such that $b\neq a$. The last diagram (\ref{sf:lgv}) contains a second order counter term that partially cancels the divergent diagrams as explained in the main text. Note that since we take $\bar{g}=0$, we suppressed all diagrams containing $\bar{g}$-vertices. Finally, we have also suppressed the diagrams whose divergences are clearly taken care of by the velocity counter terms.}
\label{fig:3rd}
\end{figure}

The computation of the counter term for $\lambda$, $\delta_\lambda$, to third order turns out to be a great challenge in bookkeeping. Figures \ref{sf:d2}-\ref{sf:d4} show all the divergent diagrams for $\bar{g}=0$ and $g=\lambda/(2\sqrt{2}\pi)$. Roughly speaking, the loops containing a propagator of both a left moving ($+$) and a right moving ($-$) field are divergent. The first two diagrams, \ref{sf:d2} and \ref{sf:d6}, are easily computed by generalizing the second order computations. The diagrams containing nested loops are slightly more subtle. Note that there are essentially two kinds: the diagrams where the inner loop is divergent (figs. \ref{sf:d8}-\ref{sf:d4}) and those where the outer loop is divergent (figs. \ref{sf:d1} and \ref{sf:d7}). The latter is still straightforward, because the integral corresponding to the inner loop can be performed without problems and the remaining integral is just that corresponding to a one loop divergent diagram. However, when the inner loop is divergent, one has to proceed with a bit more caution. We discuss this in appendix \ref{sec:3rdorder}.

Eventually, we obtain
\bea
\delta_{\lambda}=\frac{3 \lambda^3}{(4 v_F)^2} \left[\frac{2}{\varepsilon} - \log (M^2) \right].
\eea
Then working out the full CS equation for the $\lambda$-vertex 4-point function to third order, we find
\bea
\beta(\lambda)&=&M\frac{\partial}{\partial M}\delta_{\lambda} - 2 \lambda (\gamma_{\psi} +\gamma_{\chi})\nn
&=&- \frac{\lambda^3}{2 v_F^2},
\eea
where we used $\gamma_{\psi}=0$ for $g=\lambda/(2\sqrt{2}\pi)$ and $\bar{g}=0$. We thus find that $\lambda$ is marginally irrelevant and the flow is towards $\lambda=0$ as we go to lower energy scales. Note that for $u\neq v$ there will be corrections to the flow in $\lambda$, but since they will be of the form $\epsilon \mathcal{O}(\lambda^3)$, where $\epsilon$ is the velocity difference as above, these terms will be subleading for small $\epsilon$.

\begin{figure}
\includegraphics[width=\columnwidth]{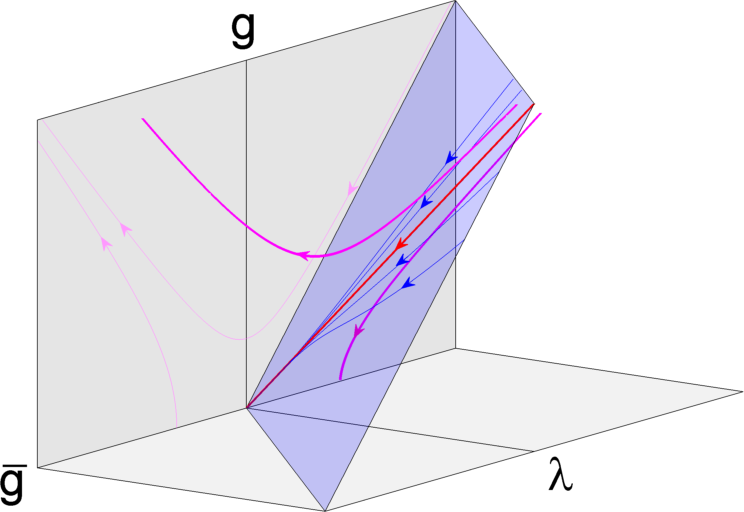}
\caption{Schematic flow diagram in the $(g,\bar{g},\lambda)$-space. The red line is the special line along which there is a slow flow towards the fixed point. The blue lines represent the flow lines in the attractive plane. In purple we show the flow for relevant perturbations towards the gapped (line above the plane) and gapless (line below the plane) phases in the $U(1)$ sector. In the $\lambda=0$-plane the flow lines are governed by the usual Kosterlitz equations. \label{fig:flow}}
\end{figure}

In appendix \ref{sec:stability} we present a careful analysis of the stability of the fixed point by considering small perturbations away from the special line. The results are summarized in the schematic flow diagram shown in figure \ref{fig:flow}. We establish that there is an attractive plane for which the flow is towards the special line and eventually terminates at the fixed point. Furthermore, this plane is a separatrix; below the plane the flow is towards $\lambda=g=0$, and $\bar{g}$ finite, while above the plane the flow is towards increasing $g$ and $\bar{g}$. This corresponds to the gapless and gapped phase in the $U(1)$ sector, respectively, that is the dashed and drawn red lines in Fig. \ref{fig:PD} (remember that $m=0$, so the Ising sector is gapless). It follows that the decoupled, Lorentz invariant fixed point with $\mathcal{N}=(3,3)$ supersymmetry can be reached upon tuning to the attractive plane and $m=0$, which requires to fine tune two parameters (as expected for the multicritical point).

\begin{table}
\begin{tabular}{|c | c | c |}

\cline{2-3}
 \multicolumn{1}{c|}{} & multicritical point & critical Ising line \\

\hline

$v_{\textrm{phys}}, u_{\textrm{phys}}$ & $v_F \ell^{-1/4} \to 0$ & $v_F e^{- a_1 \ell^{1/5}} \to 0$ \\

$\gamma$ & $(\ln \frac{1}{T})^{1/4} \to \infty$ & $ e^{a_2 (\ln \frac{1}{T})^{1/5}} \to \infty$ \\

$\Delta_m$ & $|m| (\ln \frac{1}{|m|})^{-1/4}$ & $|m| e^{- a_3 (\ln \frac{1}{|m|})^{1/5}}$ \\

\hline

\end{tabular}
\caption{We show the behavior of the physical velocities, $v_{\textrm{phys}}$ and $u_{\textrm{phys}}$, the specific heat coefficient, $\gamma$, and the Ising gap, $\Delta_m$, asymptotically close to the multicritical point (middle column) and the critical Ising line inside the gapless phase (right column). In particular, these expressions are valid for $T,|m|,1/\ell \ll 1$, where $T$ is the temperature, $m$ the Ising mass and all three parameters are measured in dimensionless units. Finally, the $a_i$ are positive constants whose values can be found in Ref. \onlinecite{Sitte}.  \label{tab:phys}}
\end{table}

Interestingly, although the Ising and $U(1)$ sectors decouple at the infra-red, the physics of the multicritical point is quite different from the usual Ising and BKT physics. This is a consequence of the slow flow towards the fixed point. In particular, we find that the correlation length in the $U(1)$ sector diverges in an unusual way as one approaches the multicritical point from the gapped phase:
\bea
\xi(\delta g_0) = \xi_0 e^{\frac{1}{4} \left( \ln \left[ \frac{\lambda_0}{\delta g_0} \right] \right)^2},
\eea
where $\xi_0$ is a constant, $\lambda_0$ is the initial value of $\lambda$ along the special line, 
and $\delta g_0>0$ is the distance from the special line in the $g$-direction (see appendix \ref{sec:corr} for details). Note that the coefficient of the logarithm squared of $1/4$ is universal. The scaling of the correlation length we obtain is a remarkable result that should be contrasted with the result for the conventional KT transition. Remember that at the KT transition the correlation length diverges as $\xi (\delta g_0) = \xi_0 \exp(c_1/\sqrt{\delta g_0})$, with $c_1$ a dimensionful non-universal constant, while for a conventional phase transition driven by a relevant operator with scaling dimension $\Delta <2$, the correlation length scales as $\xi (\delta g_0) \sim (\delta g_0)^{1/(\Delta-2)}$. The result we obtain here is a scenario in between these two well-known cases: the divergence is faster than any power law, but subexponential in the sense that $\log \xi$ grows slower than any power law; in particular, it is slower than at a traditional KT transition.

To infer the gap scaling from the correlation length we need to take into account that the $\lambda$-term gives rise to an anomalous dynamical critical exponent that depends on the RG scale. It follows that there is a relative factor of $e^{\ell (z (\ell)-1)}$ between the rescaling of the correlation length and the gap. We have to integrate this factor over the entire RG trajectory, but due to the fact that $z$ flows to $z=1$ very slowly, the result actually vanishes. This situation is also encountered in Ref. \onlinecite{Sitte} and the consequences on the physical velocities, the specific heat and the Ising gap are nicely discussed. Here the situation is very similar, the main difference is the scaling of $\lambda$. Consequently, we find $z-1 \sim \ell^{-1}$ for the dynamical critical exponent at the multicritical point, while Sitte et al. find $z-1 \sim \ell^{-4/5}$ at the Ising transition inside the gapless phase. At the multicritical point, following the arguments in Ref. \onlinecite{Sitte}, we thus find that the vanishing of the physical velocities, the specific heat coefficient divergence and the Ising gap suppression take on a different form. We summarize the results for the behavior at the multicritical point in table \ref{tab:phys} and also show the results for the critical Ising line within the gapless $U(1)$ phase for comparison.

Furthermore, for the scaling of the $U(1)$ gap we find
\bea
\Delta(\delta g_0) &=& e^{\int_0^{\ell^*} (1-z(\ell)) d\ell} \xi(\delta g_0)^{-1}\nonumber\\
&\approx& \Delta_0  \frac{1}{\sqrt{ \ln \left[ \frac{\lambda_0}{\delta g_0} \right] }} e^{-\frac{1}{4} \left( \ln \left[ \frac{\lambda_0}{\delta g_0} \right] \right)^2}. \nonumber
\eea
As with the correlation length, we see that the gap shows neither power law nor exponential scaling, but something in between. Finally, we expect that order parameter correlation functions will receive logarithmic corrections to the usual scaling due to the slow flow. To compute these corrections is beyond the scope of this paper, since one needs the field renormalizations to third order in the couplings. 

\textit{Conclusions--} we have shown that an $\mathcal{N}=(3,3)$ supersymmetry emerges in a class of lattice models. The requirements are: 1) the model has a $U(1)$ and a $\mathbb{Z}_2$ symmetry, 2) the system is tuned to the multicritical point where the Ising and BKT transitions coincide, and 3) the bare velocity of the fermionic degree of freedom is smaller than or equal to the bare velocity of the bosonic degree of freedom. Apart from the emergent extended supersymmetry, we find that the RG flow towards the fixed point is extremely slow. Consequently, the Ising-BKT multicritical point with emergent supersymmetry lies in a new universality class that is characterized, in particular, by a superpolynomial but subexponential scaling of the corrlation length in the $U(1)$ sector. It would be interesting to see if there are other multicritical points where supersymmetry emerges, in particular, also in higher dimensions and investigate if a deeper connection exists between supersymmetry and multicriticality.

\section*{Acknowledgements}
L.H. was partially funded by the John Templeton foundation and by a DOE early career grant. E.B. was supported by the Israel Science Foundation, by the Israel-US Binational Science Foundation, by the Minerva Foundation, by a Marie
Curie CIG grant, and by the Robert Rees Fund. L.H. and E.B. also acknowledge support from the NSF grant no. DMR-1103860.

\bibliographystyle{apsrev4-1}
\bibliography{refs}

\appendix

\section{Supersymmetry of the multicritical theory}\label{sec:susy}
The fixed point theory, i.e. $\mathcal{L}_0=\mathcal{L}_{\Phi}+\mathcal{L}_{\chi}$ with $u=v$ and $K=K_{\textrm{crit}}=4$, such that the cosine term is marginal, has extended supersymmetry. Extended refers to the fact that the number of supercharges, denoted by $\mathcal{N}$, is larger than one. More precisely, the fixed point theory corresponds to the 2nd minimal model in the $\mathcal{N}=2$ superconformal series with central charge, $c=3/2$. At this value of the Luttinger parameter, $K=4$, there is an additional $\mathcal{N}=(2,2)$ supersymmetry on top of the $\mathcal{N}=(1,1)$ supersymmetry, which is present for any value of $K$. The notation $\mathcal{N}=(\mathcal{N}_L,\mathcal{N}_R)$ implies that there are $\mathcal{N}_{L(R)}$ supercharges in the left (right) moving sector. 

The supercharges are fermionic operators and as a consequence they generate transformations that map bosonic operators into fermionic operators and vice versa. Furthermore, the supercurrents form a a closed superalgebra, an algebra with both commutators and anti-commutators, with the stress-energy tensor and, in the case of extended supersymmetry, a $U(1)$ current \cite{DiVecchia85,Boucher86}.

The $\mathcal{N}=(1,1)$ supersymmetry is generated by the supercurrents $G_{L}^0=\chi_L \partial_z \Phi_L$ and $G_{R}^0=\chi_R \partial_{\bar{z}} \Phi_R$, where $z=x + \imath v \tau$ and $\Phi(z,\bar{z})=\Phi_L(z)+\Phi_R(\bar{z})$. We define the infinitesimal transformations generated by the conserved charges associated to these supercurrents for some field $\mathcal{O}$ in the left moving sector as $\mathcal{O} \to \mathcal{O} + \xi \delta_{0,L} \mathcal{O}$, where $\xi$ is a Grassmannian variable, and similarly for the right moving sector. In particular, we obtain the following transformations:
\bea
\left\{ \begin{array}{lll} \delta_{0,L} \chi_L & = & -\frac{1}{2}\partial_z \Phi_L \\
\delta_{0,L} \partial_z \Phi_L & = & \partial_z \chi_L \end{array} \right. \, \textrm{and} \,
\left\{ \begin{array}{lll} \delta_{0,R} \chi_R & = & -\frac{1}{2}\partial_{\bar{z}} \Phi_R \\
\delta_{0,R} \partial_{\bar{z}} \Phi_R & = & \partial_{\bar{z}} \chi_R \end{array} \right.  .\nonumber
\eea
The theory is invariant under these transformations that map the boson into the fermion and vice versa.

The $\mathcal{N}=(2,2)$ supersymmetry at $K=4$ is generated by the supercharges $G^{\pm}_L=\chi_L \exp(\pm \imath \sqrt{2} \Phi_L)$ for the left movers and similarly for the right movers. The corresponding infinitesimal transformations for the fields in the left moving sector are
\bea
\left\{ \begin{array}{lll} \delta_{\pm,L} \chi_L & = & -\frac{1}{2}e^{\pm \imath \sqrt{2} \Phi_L} \\
\delta_{\pm,L} \partial_z \Phi_L & = & \mp \imath \sqrt{2} \chi_L e^{\pm \imath \sqrt{2} \Phi_L} \end{array} \right. , \nonumber
\eea
and similarly for the right moving sector. 

Finally, in Ref. \onlinecite{Bauer13} we established that the interacting theory preserves an $\mathcal{N}=2$ supersymmetry for $u=v$, $m=0$ and $\lambda=2 \sqrt{2} \pi g$. This corresponds precisely to the line of fixed points to second order in the couplings. Note that this is an extended supersymmetry in a theory that is not Lorentz invariant due to the presence of the $\lambda$-term, as a consequence the supersymmetry tranformations mix the left and right moving sectors. The supercurrent conservation relation reads  
\bea
\bar{\partial} \tilde{G}_L^{\pm}+\partial \tilde{G}_R^{\pm}=0, \nonumber
\eea
with
\bea
\tilde{G}_L^{\pm} &=& G_L^{\pm} + \frac{g}{2} \psi_L e^{\mp \imath \sqrt{2} \Phi_R}  \nonumber\\
\tilde{G}_R^{\pm}&=& \mp \imath G_R^{\pm} \mp \imath \frac{g}{2} \psi_R e^{\pm \imath \sqrt{2} \Phi_L} . \nonumber
\eea

\section{Fermionization of the theory}\label{sec:fermionize}
To fermionize the boson we introduce a second, redundant bosonic field, with velocity $v_s$ and Luttinger parameter $K_s$, which is free and completely decoupled from the other fields. The fermionized theory is then found by retracing the bosonization steps. Consider the following Hamiltonian for four types of complex fermions, $\psi_{ap}$, with $a=1,2$ and $p=\pm$,
\bea
H_F&=&H_0 + \bar{g} \sum_{\substack{a=1,2 \\ b=1,2}} \psi_{a+}^{\dagger} \psi_{a+}  \psi_{b-}^{\dagger} \psi_{b-} \nonumber\\
& & + 2 g_v \sum_{p=\pm}  \psi_{1p}^{\dagger} \psi_{1p}  \psi_{2p}^{\dagger} \psi_{2p}. \nonumber
\eea
The first term, $H_0$, is the standard free fermion part of the Hamiltonian for all four types of fermions with velocity $v_F$. The latter two terms are interaction terms, the $\bar{g}$-term is often called a dispersion term scattering left onto right fermions, the $g_v$-term is a forward scattering term. We bosonize this Hamiltonian using
\bea
\psi_{ap}=\frac{1}{\sqrt{2 \pi}} e^{\imath \varphi_{ap}}, &\quad& \psi_{ap}^\dagger \psi_{ap} = \frac{1}{2 \pi}  \partial_x \varphi_{ap}, \nonumber\\
\varphi_{ap}& =& \phi_a + p \ \theta_a, \nonumber\\
\Phi_c = \frac{1}{\sqrt{2}}(\phi_1 +\phi_2), &\quad& \Phi_s = \frac{1}{\sqrt{2}}(\phi_1 - \phi_2), \nonumber\\
\Theta_c= \frac{1}{\sqrt{2}}(\theta_1 +\theta_2), &\quad& \Theta_s = \frac{1}{\sqrt{2}}(\theta_1 - \theta_2), \nonumber
\eea
where $\theta_a$ is the so-called dual boson, defined by $\partial_x \theta_a = - \Pi_a$, where $\Pi_a$ is the canonical conjugate of the field $\phi_a$. With these definitions we obtain
\bea
H_B &=& \frac{v_c}{2\pi} \left[ K_c (\partial_x \Theta_c)^2 +\frac{1}{K_c} (\partial_x \Phi_c)^2\right] \nonumber\\
& & + \frac{v_s}{2\pi} \left[ K_s (\partial_x \Theta_s)^2 +\frac{1}{K_s} (\partial_x \Phi_s)^2\right], \nonumber
\eea
with
\bea
v_c &=& v_F \sqrt{\left(1+\frac{g_v}{\pi v_F}\right)^2 - \left(\frac{\bar{g}}{\pi v_F}\right)^2} = v_F +  \frac{g_v}{\pi }+ \dots, \nonumber\\
K_c &=& \sqrt{\frac{1+\frac{g_v}{\pi v_F} - \frac{\bar{g}}{\pi v_F}}{1+\frac{g_v}{\pi v_F} + \frac{\bar{g}}{\pi v_F}}} = 1-  \frac{\bar{g}}{\pi v_F}+ \dots, \nonumber\\
v_s &=&  v_F -  \frac{g_v}{\pi }+ \dots, \quad K_s= 1 . \nonumber
\eea
It follows that the fermionic Hamiltonian above allows us to recover the free part of our bosonic theory. Furthermore, we find that the cosine-term can be recovered from an umklapp-scattering term for the fermions,
\bea
2 \pi^2 \left( \psi_{1+}^{\dagger} \psi_{1-}  \psi_{2+}^{\dagger} \psi_{2-} +\textrm{h.c.} \right) = \cos(2 \sqrt{2} \Phi_c),  \nonumber
\eea
Comparing this with the cosine-term in our bosonic action implies that we should identify $\Phi \equiv 2 \Phi_c$. If we also take $\Theta \equiv \Theta_c/2$ to ensure the canonical commutation relation between $\Phi$ and $\Theta$, we find 
\bea
v=v_c=  v_F +  \frac{g_v}{\pi}, \, K=4 K_c= 4 -  \frac{4 \bar{g}}{\pi v_F}. \nonumber
\eea
Finally, the $\lambda$-term fermionizes to
\bea
 -\lambda (\partial_x \Phi) \imath \chi_R \chi_L =  - \sqrt{2} \pi \lambda  \sum_{\substack{p=\pm \\ a=1,2}} \left( \psi_{a,p}^{\dagger}\psi_{a,p} \right) \imath \chi_R \chi_L . \nonumber
\eea

For completeness we checked that for $u+\epsilon_u=v_F+\epsilon_v +g_v/\pi$, corresponding to $u=v$, and $m=0$ the fermionic theory has $\mathcal{N}=(1,1)$ supersymmetry. The fermionized supercharges read: $\chi_L (\psi_{1-}^{\dagger} \psi_{1-} + \psi_{2-}^{\dagger} \psi_{2-})$ and $\chi_R (\psi_{1+}^{\dagger} \psi_{1+} + \psi_{2+}^{\dagger} \psi_{2+})$. Additionally, fixing also $\lambda=g=\bar{g}=0$ there is an $\mathcal{N}=(2,2)$ supersymmetry generated by the supercharges $\chi_L \psi_{1-} \psi_{2-}$ and $\chi_L \psi_{2-}^{\dagger} \psi_{1-}^{\dagger}$ in the left-moving sector and similarly for the right-moving sector.

\section{RG analysis to second order for fermionized theory}\label{sec:2ndorder}

\begin{figure}
\centering
\begin{subfigure}{.3\columnwidth}
  \centering
  \includegraphics[width=.9\columnwidth]{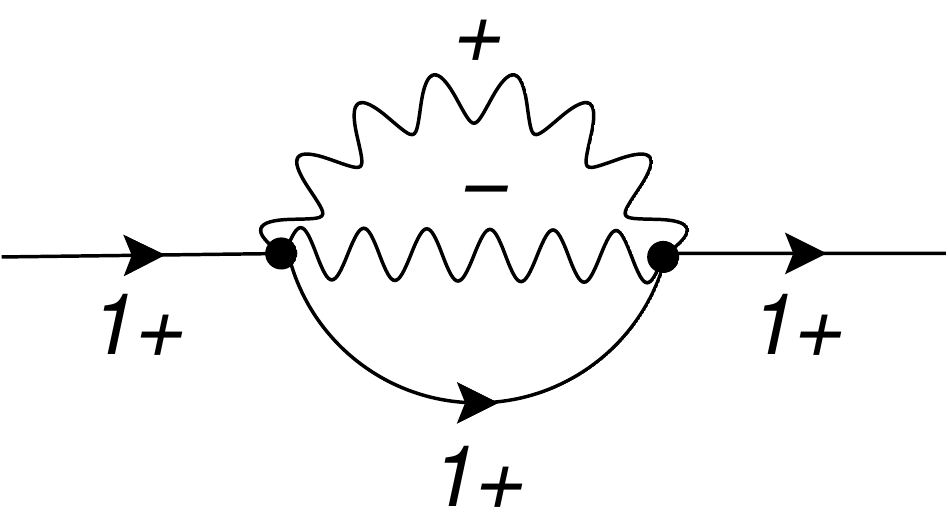}
  \caption{$\lambda^2$}
\end{subfigure}%
\begin{subfigure}{.3\columnwidth}
  \centering
  \includegraphics[width=.9\columnwidth]{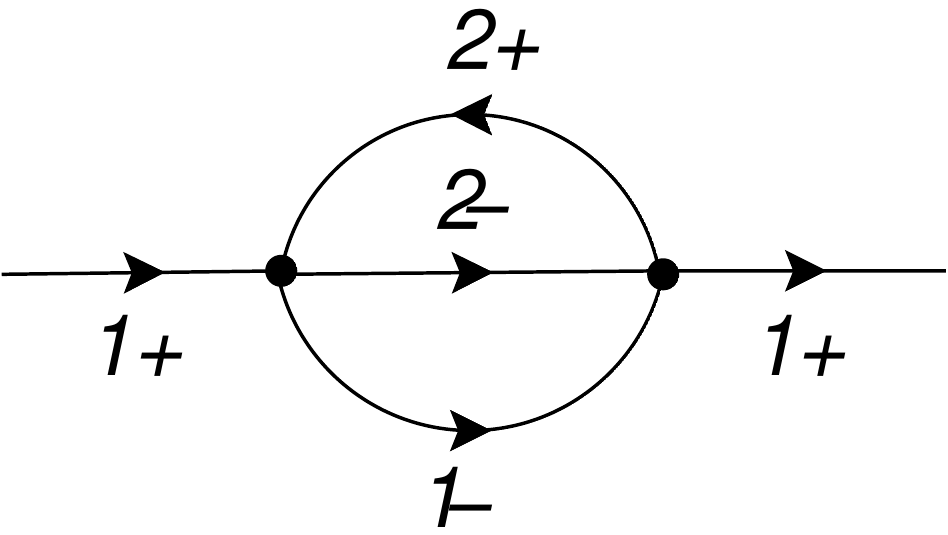}
  \caption{$g^2$}
\end{subfigure}%
\begin{subfigure}{.3\columnwidth}
  \centering
  \includegraphics[width=.9\columnwidth]{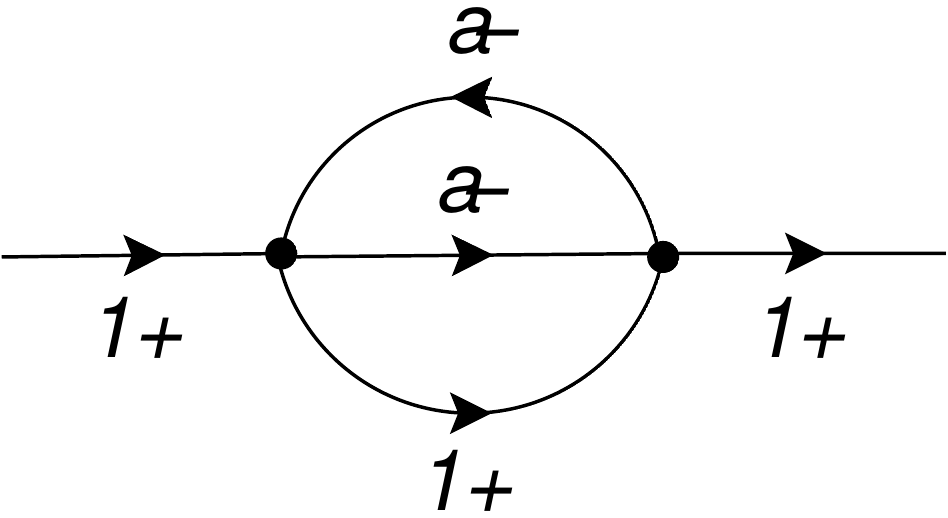}
  \caption{$\bar{g}^2$}
\end{subfigure}
\caption{Divergent diagrams in the complex fermion propagator to second order in the couplings. Straight (wiggly) lines represent complex (real) fermion propagators, respectively, and $a=1,2$.}
\label{fig:fermprop}
\end{figure}

\begin{figure}
\centering
\begin{subfigure}{.3\columnwidth}
  \centering
  \includegraphics[width=.9\columnwidth]{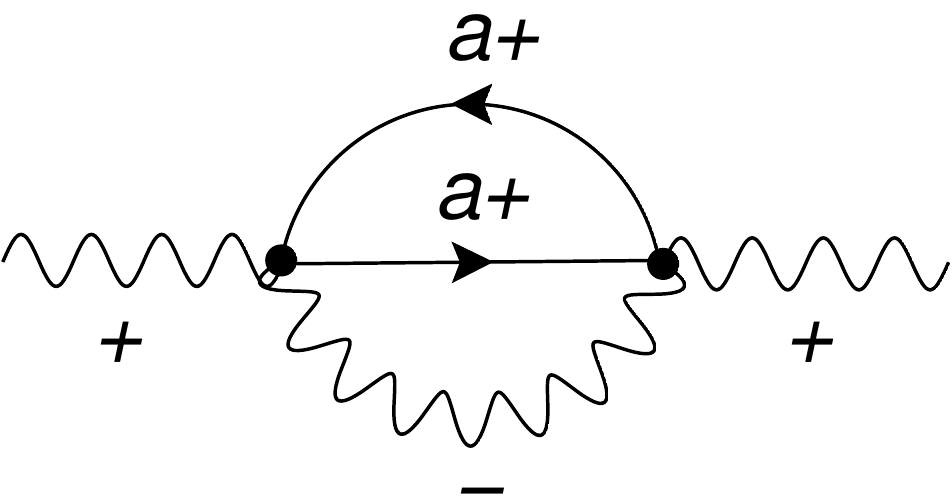}
  \caption{$\lambda^2$ \label{sf:prop}}
\end{subfigure}%
\begin{subfigure}{.3\columnwidth}
  \centering
  \includegraphics[width=.9\columnwidth]{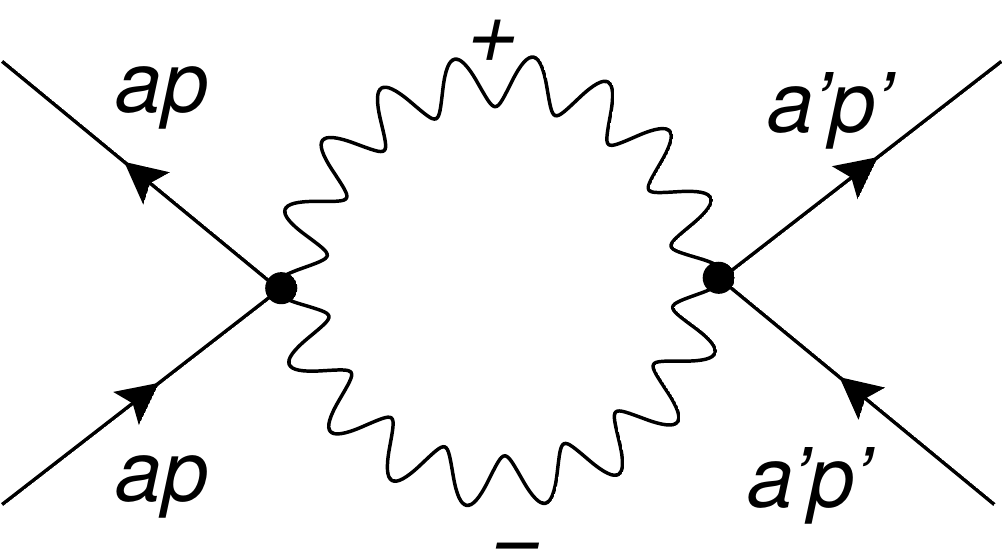}
  \caption{$\lambda^2$ \label{sf:gvgbarL2}}
\end{subfigure}%
\begin{subfigure}{.3\columnwidth}
  \centering
  \includegraphics[width=.9\columnwidth]{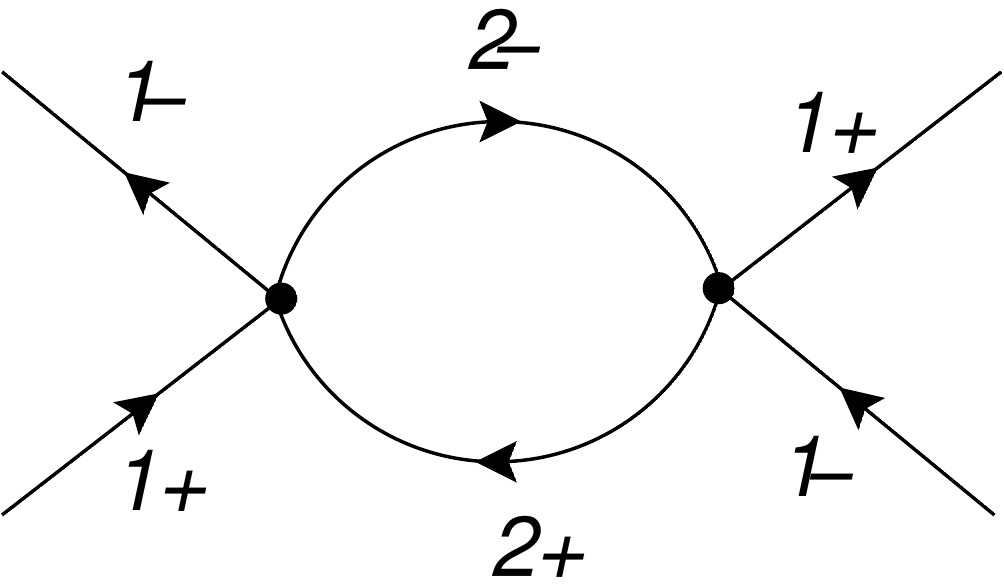}
  \caption{$g^2$ \label{sf:gbarg2}}
\end{subfigure} \\
\begin{subfigure}{.3\columnwidth}
  \centering
  \includegraphics[width=.9\columnwidth]{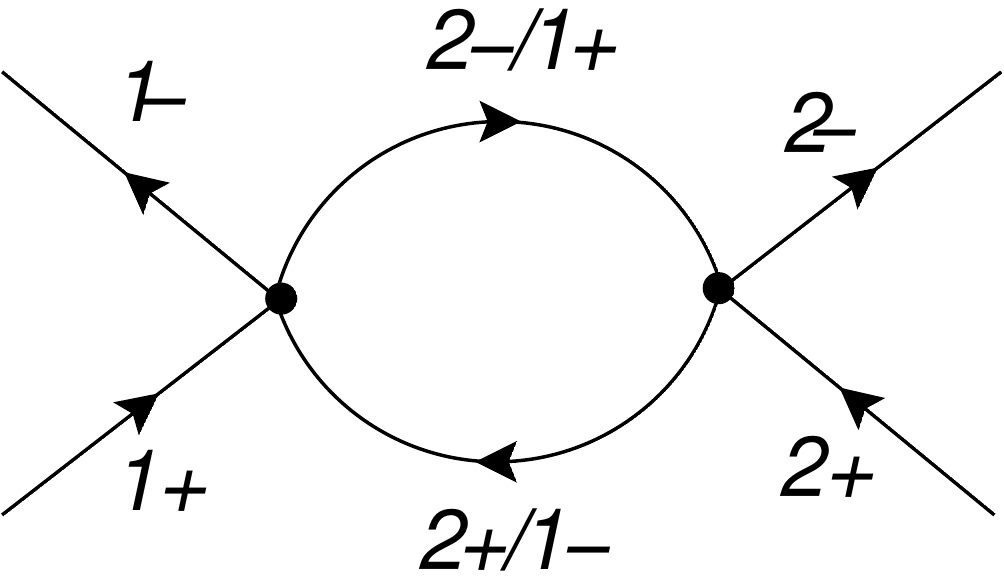}
  \caption{$g \bar{g}$/$\bar{g} g$  \label{sf:gAggbar}}
\end{subfigure}
\begin{subfigure}{.3\columnwidth}
  \centering
  \includegraphics[width=.9\columnwidth]{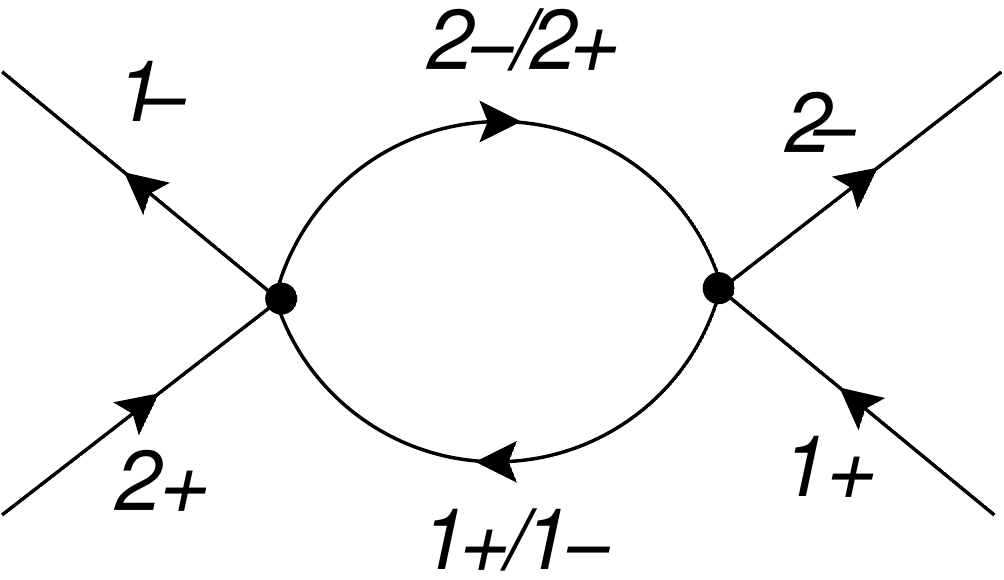}
  \caption{$g \bar{g}$/$\bar{g} g$ \label{sf:gBggbar}}
\end{subfigure}\\
\begin{subfigure}{.5\columnwidth}
  \centering
  \includegraphics[width=.9\columnwidth]{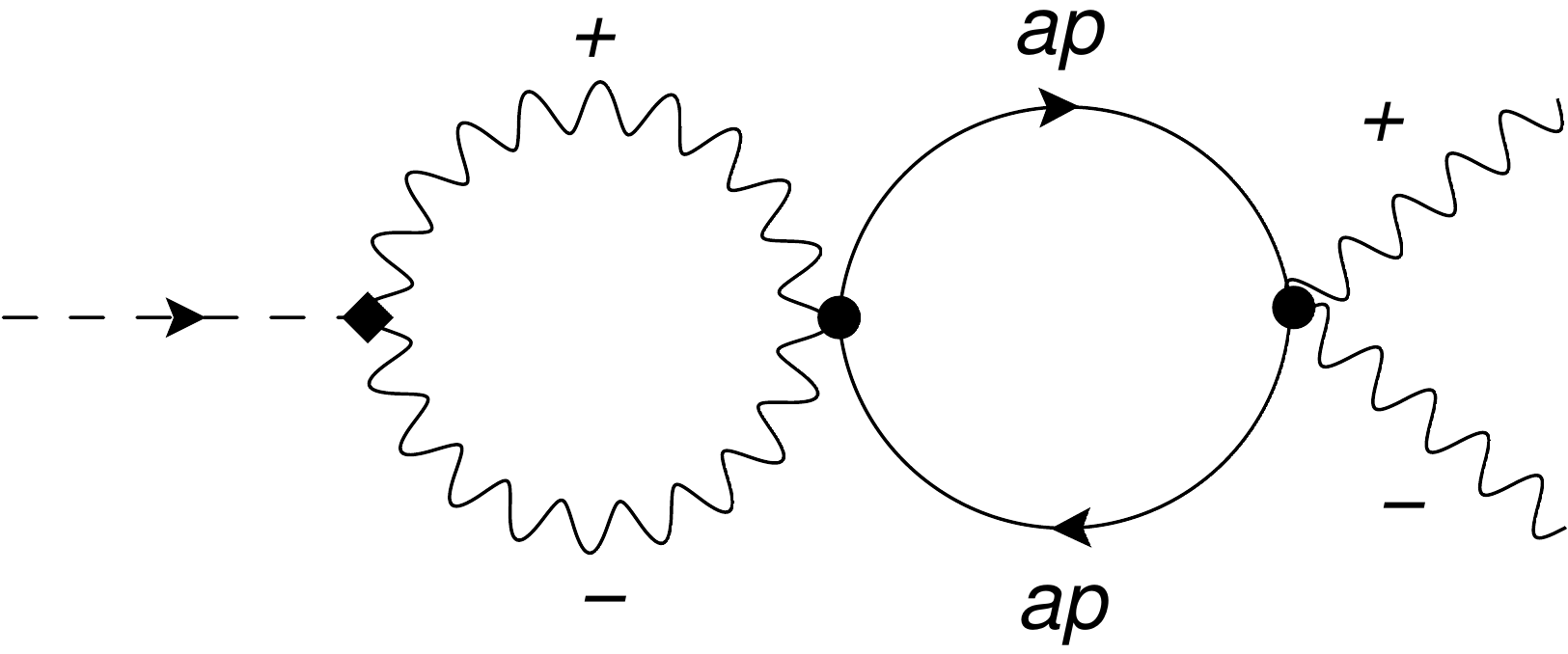}
  \caption{$\lambda^2$ \label{sf:mass}}
\end{subfigure}
\caption{Divergent diagrams in the real fermion propagator (\ref{sf:prop}) and in the vertex 4-point functions to second order in the couplings. The $\lambda^2$-diagram (\ref{sf:gvgbarL2}), where $a,a'=1,2$ and $p,p'=\pm$, contributes to $\delta_{g_v}$ for $a\neq a'$ and $p=p'$, and to $\delta_{\bar{g}}$ for $p\neq p'$. We note that for $a=a'$ and $p=p'$ the diagram is zero due to fermion sign cancellations of all possible ways to contract the complex fermions. The $g^2$-diagram (\ref{sf:gbarg2}) also contributes to $\delta_{\bar{g}}$. The $g\bar{g}$/$\bar{g}g$-diagrams, (\ref{sf:gAggbar}) and (\ref{sf:gBggbar}), contribute to $\delta_{g}$. In these diagrams the left vertex is $g$ and the right vertex is $\bar{g}$ for one choice of the propagators in the loop, namely the first choice, and vice versa for the second choice. Finally, the $\lambda^2$-diagram (\ref{sf:mass}), where $a=1,2$ and $p=\pm$, contributes to the anomalous dimension function, $\gamma_m$, of the real fermion mass-term operator, $\mathcal{O}_m$ (represented by the dashed line, the diamond and the two real fermion propagators).}
\label{fig:vertex}
\end{figure}

In this section we compute the flow equations for the fermionized theory by computing the beta functions, $\beta (g_i )=-\partial g_i/\partial (\ln M)$, for the couplings $g_i$, to second order. Schematically, we first perturbatively compute $n$-point functions in the presence of the interaction terms. We then introduce counter terms for the divergent diagrams by imposing the renormalization conditions at the renormalization scale $M$. Finally, we use the fact that the $n$-point functions obey the Callan-Symanzik (CS) equation to obtain the field renormalizations and the beta functions. For an $n$-point function with $n_{\psi}$ and $n_{\chi}$ external legs for the complex and real fermions respectively, the CS equation reads
\bea
& & \Big[ M \frac{\partial}{\partial M} - \sum_i \beta(g_i) \frac{\partial}{\partial g_i}  \nonumber\\
& & \quad + n_{\psi} \gamma_{\psi} + n_{\chi} \gamma_{\chi} \Big] G^{(n)}(\{\Omega_m,q_m \})=0, \nonumber
\eea  
where $g_i$ are all the couplings, $n=n_{\psi}+n_{\chi}$ and $m$ runs over all $n$ external legs. Note that we have defined the beta functions with the sign convention commonly used in the condensed matter community (see e.g. Shankar \cite{Shankar}), which is the opposite sign convention to that typically used in field theory textbooks (see e.g. Peskin-Schroeder \cite{Peskin}). The field renormalizations, $\gamma_{\psi,\chi}$, can be determined from the CS equation for the fermion propagators. In fig. \ref{fig:fermprop} we show the divergent diagrams that contribute to the complex fermion propagator, $G^{(2)}_{1+} (\Omega,q)=\langle \psi^{\dagger}_{1+}(\Omega,q)\psi_{1+}(\Omega,q)\rangle$, to second order. Roughly speaking the loops containing a propagator of both a left-moving (+) and a right-moving (-) field are divergent. To compute these diagrams we set $u=v_F$ in the free action \footnote{Note that this simplification is needed, because in the fermionized language already at second order in the couplings, the divergent terms come from two-loop diagrams (see Fig. \ref{fig:fermprop}), while in the bosonic language these were one-loop diagrams.}, which corresponds to $u=v$ in the bosonic case. We are primarily interested in computing the leading order flow of $\lambda$ and we will find that even for $u=v$ this is non-zero at third order in the couplings. For $u\neq v$ there will be corrections to this, but since they will be of the form $\epsilon \mathcal{O}(\lambda^3)$, where $\epsilon$ is the velocity difference as above, these terms will be subleading for small $\epsilon$. We can thus content with setting $u=v$ here. The divergence corresponding to the two loop diagram can then be computed using the integrals given in appendix \ref{app:int}. Note that since the $\lambda$-term breaks Lorentz invariance, we need to introduce separate counter terms for the frequency and momentum terms in the free action, i.e. in momentum space the counter terms read
\bea
\sum_{a=1,2} \sum_{p=\pm} (-\imath \Omega \delta_{Z_{\psi},0}+p vq\delta_{Z_{\psi},1})\psi_{a,p}^{\dagger}\psi_{a,p}.
\eea
Imposing the renormalization conditions we obtain
\bea
\delta_{Z_{\psi},0} & =&\Big(\frac{\bar{g}^{2}+2\pi^4 g^{2}}{8\pi^2 v_F^{2}}-\frac{\lambda^{2}}{32 v_F^{2}}\Big)\Big[\frac{2}{\varepsilon}-\log(M^{2})\Big], \nonumber\\
\delta_{Z_{\psi},1} & =&\Big(\frac{\bar{g}^{2}+2\pi^4 g^{2}}{8\pi^2 v_F^{2}}+\frac{\lambda^{2}}{32 v_F^{2}}\Big)\Big[\frac{2}{\varepsilon}-\log(M^{2})\Big], \nonumber
\eea
where $\varepsilon$ is the dimensional regularization, $d=2-\varepsilon$, and $M$ the renormalization scale. We can now solve the CS equation to find
\bea
\gamma_{f} & =&- \frac{\bar{g}^{2}+2\pi^4 g^{2}}{8\pi^2 v_F^{2}}+\frac{\lambda^{2}}{32 v_F^{2}} , \nn
\beta(\epsilon_{v}) & =& -\frac{\lambda^{2}}{8 v_F}. \nonumber
\eea
For the real fermion propagator the divergent diagram is shown in fig. \ref{sf:prop}. This leads to
\bea
\gamma_{\chi} & =&\frac{\lambda^{2}}{(4 v_F)^{2}},\nn
\beta(\epsilon_{u}) & =&-\frac{\lambda^{2}}{4 v_F}. \nonumber
\eea
The next step is to compute the counter terms for the interaction terms, $g_v$, $\bar{g}$, $g$ and $\lambda$. The divergent diagrams in the relevant 4-point functions are given in fig. \ref{fig:vertex}. We find
\bea
\beta(g_{v}) &=& -\frac{\pi \lambda^{2}}{8 v_F}, \nn
\beta(\bar{g})&=& - \frac{\pi}{4 v_F} \left( \lambda^{2}-8 \pi^2 g^{2} \right), \nn
\beta(g) &=& \frac{2 g\bar{g}}{\pi v_F},\nn
\beta(\lambda) &=& 0 .
\eea

Finally, obtaining the beta function for the mass-term is slightly more subtle. One finds that $\beta(m)=(2-d_m-\gamma_m)m$, where $d_m=1$ is the scaling dimension of the operator $\mathcal{O}_m= \imath \chi_R \chi_L$ and $\gamma_m$ is the anomalous dimension function for this operator. We find $\gamma_m$ from the CS equation for the Green's function containing the local operator $\mathcal{O}_m$:
\bea
\left[M\frac{\partial}{\partial M}-\sum_i \beta(g_i) \frac{\partial}{\partial g_i}+2 \gamma_{\chi} +\gamma_m \right] G^{(2;1)}=0,
\eea 
where $G^{(2;1)}=\langle \chi_R (\omega_1,p_1) \chi_L (\omega_2,p_2) \mathcal{O}_m (\omega,p)\rangle$ and $g_i$ runs over all couplings except the mass. We find $\gamma_m=\lambda^2/(8 v_F^2)$ and thus,
\bea
\beta(m)=m\left(1-\frac{\lambda^2}{8v_F^2}\right).
\eea
To see that these results agree with the RG equations obtained from real space RG we need to restrict the comparison to the case $u=v$. We confirm that as before the velocities remain equal: $\beta(v)=\beta(\epsilon_v)+\beta(g_v)/\pi= \pi \lambda^2/(4 v_F)$, which equals $\beta(\epsilon_u)$. Furthermore, we easily check that the velocity, $v_s$, of the decoupled additional boson that we introduced to fermionize theory, indeed does not flow: $\beta(v_s)=\beta(\epsilon_v)-\beta(g_v)/\pi=0$. Finally, one easily verifies the other equations using the relation between $K$ and $\bar{g}$.

\section{Details on RG analysis to third order}\label{sec:3rdorder}

In this section we provide some details on the computation of the third order term of the beta function for $\lambda$ on the special line. The main challenge is to compute the counter term for $\lambda$ to third order. Figures \ref{sf:d2}-\ref{sf:d4} show all the divergent diagrams for $\bar{g}=0$ and $g=\lambda/(2\sqrt{2}\pi)$ that contribute to the counter term. The first two diagrams, \ref{sf:d2} and \ref{sf:d6}, are easily computed by generalizing the second order computations. The diagrams containing nested loops are slightly more subtle. Note that there are essentially two kinds: the diagrams where the inner loop is divergent (figs. \ref{sf:d8}-\ref{sf:d4}) and those where the outer loop is divergent (figs. \ref{sf:d1} and \ref{sf:d7}). The latter is still straightforward, because the integral corresponding to the inner loop can be performed without problems and the remaining integral is just that corresponding to a one loop divergent diagram. However, when the inner loop is divergent, one has to proceed with a bit more caution.

In the diagrams where the inner loop is divergent one encounters the following type of integral
\bea
\int\frac{ d\omega dk}{(2\pi)^{2}}\frac{1}{\bar{z}(\bar{z}+\bar{w}_{3}+\bar{w}_{4})}\Big(\frac{2}{\epsilon}-\log(|z-w_{2}|^{2})+\dots\Big), \nonumber
\eea
where $z=\imath \omega+vk$. Although we cannot really compute this integral, we can find the logarithmically
divergent part. This is the part that will contribute in the CS equation
via the counter term. To find the logarithmically divergent part of
the last integral we proceed as follows. First we define
\bea
I_{4}(M)=\int_{z}\frac{1}{\bar{z}(\bar{z}+\bar{z}_{0})}\log(|z-M|^{2}), \nonumber
\eea
where we used $\int_{z}\equiv\int\frac{d\omega dk}{(2\pi)^{2}}$ and we
have chosen $w_{2}=M$, without loss of generality. It follows that

\begin{widetext}
\bea
\frac{\partial}{\partial M}I_{4}(M) & =&\int_{z}\frac{1}{\bar{z}(\bar{z}+\bar{z}_{0})}\Big(\frac{-1}{z-M}+\frac{-1}{\bar{z}-M}\Big)\nn
 & =&\Gamma(3)\Big(\prod_{i=1}^{3}\int_{0}^{1}dx_{i}\Big)\int_{z}\frac{-\delta(1-\sum_{i=1}^{3}x_{i})(z-z_{0})(\bar{z}-M)z}{[x_{1}|z+z_{0}|^{2}+x_{2}|z-M|^{2}+x_{3}|z|^{2}]^{3}}+[(z,\bar{z}-M)\longleftrightarrow(\bar{z},z-M)]\nn
 & =&-2\Big(\prod_{i=1}^{3}\int_{0}^{1}dx_{i}\Big)\delta(1-\sum_{i=1}^{3}x_{i}) \int_{\ell}\frac{(\ell+(1-x_{1})z_{0}+x_{2}M)(\bar{\ell}-x_{1}\bar{z}_{0}-(1-x_{2})M)(\ell-x_{1}z_{0}+x_{2}M)}{[|\ell|^{2}+\Delta]^{3}}\nn
 & &+[(-(1-x_{2})M)\longleftrightarrow(x_{2}M)],\nonumber
\eea
\end{widetext}
where we in the last step we completed the square in the denominator,
changed integration variable, $\ell=z+x_{1}z_{0}-x_{2}M$, and defined
$\Delta=x_{1}|z_{0}|^{2}+x_{2}M^{2}-|x_{1}z_{0}-x_{2}M|^{2}$. The
divergent part of this integral has the numerator constant. Furthermore,
we now choose $z_{0}=0$ without loss of generality as we will argue
shortly. It follows that $\Delta$ simplifies to $\Delta=x_{2}(1-x_{2})M^{2}$
and the constant part of the numerator becomes $x_{2}^{2}(1-x_{2})M^{3}$. We then find
\bea
\frac{\partial}{\partial M}I_{4}(M) & =& 4 \frac{1}{8\pi v}\int_{0}^{1}dx_{1}\int_{0}^{1-x_{1}}dx_{2}\frac{x_{2}^{2}(1-x_{2})M^{3}}{[x_{2}(1-x_{2})M^{2}]^{2}}\nn
& &+\dots\nn
 & =&-\frac{1}{2\pi v}\frac{1}{M}+\dots \nonumber
\eea
Thus we obtain that the logarithmically divergent part of $I_{4}(M)$
reads $-\log M/(2\pi v)$. Now let us briefly comment on our choice
of setting $z_{0}$ to zero. What one might worry about is that the
coefficient of the $\log M$ term depends on this choice, or more
precisely on the ratio of $z_{0}$ and $z_{3}$. This would
imply, however, that $I_{4}(M)$ diverges as $z_{3}=M\to0$, while
$z_{0}$ remains finite. By inspecting $I_{4}(M)$ it is clear that
it only diverges as both $M$ and $z_{0}$ vanish at the same time.
Therefore, we can exclude this possibility and find that setting $z_{0}=0$,
gives us the right coefficient of the $\log M$ term. 

At this point, we know how to compute all the relevant diagrams and the remaining challenge is to get all the signs and factors of 2 correct. A good check on the computation is provided by the fact that some of the divergences at third order are canceled by second order counter terms. For instance, the last diagram, fig. \ref{sf:lgv}, contains the second order counter term, $\delta_{g_v}$, and should cancel the divergence of the diagram in fig. \ref{sf:d2} with $ap=2+$. We verified that this is indeed the case. Furthermore, the divergences of the diagram in fig. \ref{sf:d2} with $a=1,2$ and $p=-$ and of the diagram in fig. \ref{sf:d4} should be canceled by a diagram containing the counter term for $\bar{g}$. However, since $\delta_{\bar{g}}=0$ on the line of second order fixed points, we find as expected that these diagrams cancel for $g=\lambda/(2\sqrt{2}\pi)$. Finally, we can argue that the divergences of the diagrams in fig. \ref{sf:d2} with $ap=1+$ and in fig. \ref{sf:d3} and the diagrams in fig. \ref{sf:d6} and fig. \ref{sf:d8} should cancel pairwise due to fermion statistics. The diagrams can be transformed into one another by exchanging two identical fermionic fields ($\psi_{1+}^{\dagger}$), which leads to a relative minus sign. We find that this indeed works out.

Eventually, we are left with two divergent diagrams: \ref{sf:d1} and \ref{sf:d7}. From these we find
\bea
\delta_{\lambda}=\frac{3 \lambda^3}{(4 v_F)^2} \left[\frac{2}{\varepsilon} - \log (M^2) \right].
\eea 
Then working out the full CS equation for the $\lambda$-vertex 4-point function to third order, we find
\bea
\beta(\lambda)&=&M\frac{\partial}{\partial M}\delta_{\lambda} - 2 \lambda (\gamma_{\psi} +\gamma_{\chi})\nn
&=&- \frac{\lambda^3}{2 v_F^2},
\eea
where we used $\gamma_{\psi}=0$ for $g=\lambda/(2\sqrt{2}\pi)$ and $\bar{g}=0$. We thus find that the beta function is negative for positive $\lambda$ and positive for negative $\lambda$, which means that $\lambda$ is marginally irrelevant and the flow is towards $\lambda=0$ as we go to lower energy scales.

\section{Stability of the fixed point}\label{sec:stability}
We have computed the flow of $\lambda$ on the special line $\bar{g}=0$ and $g=\lambda/(2\sqrt{2}\pi)$ to third order in the couplings and found that the flow along this line is towards the decoupled fixed point: $\bar{g}=g=\lambda=0$. In this section we show that close to the fixed point there is an attractive plane in the $(\lambda,g,\bar{g})$-parameter space for which the flow is towards this special line and eventually towards the fixed point. This established the stability of the decoupled fixed point.

We first determine all the flow equations to third order in the couplings on the special line. On this line we have: 1) $\lambda/(2\sqrt{2}\pi)-g=0+\mathcal{O}(g_i g_j)$, 2) $\bar{g}=0+\mathcal{O}(g_i g_j)$, and 3) $\beta(g_i)=0+\mathcal{O}(g_i g_j g_k)$, where $g_i=g,\lambda,\bar{g}$. From 1) it follows that $\beta(\lambda)/(2\sqrt{2}\pi)=\beta(g)+\mathcal{O}(g_i \beta(g_i))$, finally, using 3) we find $\beta(\lambda)/(2\sqrt{2}\pi)=\beta(g)+\mathcal{O}(g_i g_j g_k g_l)$. That is, to third order in the couplings the beta function of $g$ on the special line is completely determined by the beta function of $\lambda$. Similarly, we easily obtain that the beta function of $\bar{g}$ vanished to third order in the couplings on the special line. Solving these flow equations, we obtain
\bea
\lambda(\ell)=2 \sqrt{2} \pi g(\ell)=\frac{\lambda_0 v_F}{\sqrt{\lambda_0^2 \ell + v_F^2}}, \quad \bar{g}(\ell)=0, \nonumber
\eea
where $\ell=-\ln M$ and $\lambda_0=\lambda(0)$ is a (small) constant. To determine the flow close to the special line, we add a small perturbation to this solution. We write the deviations as $\delta \lambda, \delta g$ and $\delta \bar{g}$ and plug the perturbed solution back into the full flow equations. Keeping only second order terms in the couplings and in the deviations, we obtain 
\bea
\beta(\delta \bar{g})&=&-\frac{\pi}{4 v_F}(2 \lambda \delta \lambda + \delta \lambda^2 -4 \sqrt{2}\pi \lambda \delta g - 8 \pi^2 \delta g^2) \nonumber\\
\beta(\delta g)&=& \frac{2}{\pi v_F}\delta \bar{g}(\lambda/(2\sqrt{2}\pi)+\delta g) \nonumber\\
\beta(\delta \lambda)&=&0 \nonumber
\eea
We can linearize and solve these equations in the regime $\delta \lambda,\delta g, \delta \bar{g} \ll g, \lambda$. Since $\beta(\delta \lambda)=0$, we find that $\delta \lambda$ is constant. This constant can easily be absorbed into $\lambda(\ell)$, so we can set $\delta \lambda=0$ without loss of generality. The linearized equations for this choice read
\bea
\frac{d}{dx}\delta \bar{g}&=&\frac{ \sqrt{2}\pi^2}{v_F} \delta g \nonumber\\
\frac{d}{dx}\delta g&=& \frac{1}{ \sqrt{2} \pi^2 v_F} \delta \bar{g} \nonumber
\eea
where $\frac{d}{dx}\equiv \frac{1}{\lambda(\ell)} \frac{d}{d\ell}$. The general solution to these linearized equations is 
\beq
\left( \begin{array}{c} \delta \bar{g} \\ \delta g \end{array} \right) = \vec{v}_+ e^{x/v_F} + \vec{v}_- e^{-x/v_F} , \label{flowpert}
\eeq
with $\vec{v}_{\pm}=(\sqrt{2}\pi^2, \pm 1)$. Finally, $x$ can be found by integrating its defining equation. This results in
\beq
x(\ell)=\frac{2v_F}{\lambda_0}\sqrt{\lambda_0^2 \ell + v_F^2}-\frac{2v_F^2}{\lambda_0}.
\eeq
It follows that the deviations from the special line flow to zero if we move away in the $\vec{v}_-$ direction. We thus remain in the regime of validity of our approximation: $\delta \lambda,\delta g, \delta \bar{g} \ll g, \lambda$. We conclude that there is an attractive direction at every point on the special line and thus a plane in the $(\lambda,g,\bar{g})$-parameter space for which the flow is towards and along this special line, eventually terminating at the fixed point.

Finally, we mention that deviations from the special line that are not parallel to $v_-$, take us either above or below the attractive plane. In both cases the deviations from the line start to grow and at some point the regime of validity of the linearized flow equations around the special line breaks down. However, since $\lambda$ is decreasing under the flow, while $g$ and/or $\bar{g}$ are increasing, we eventually reach a regime where the flow equations of the decoupled fixed point are valid and thus $g$ and $\bar{g}$ continue to flow according to the usual Kosterlitz equations. In particular, they continue to grow if the flow started above the attractive plane, whereas $g$ flows to zero if we started below the attractive plane. This corresponds to the gapped and gapless phase in U(1) sector, respectively (remember that the Ising sector is always gapless). We have checked these predictions by solving the full flow equations numerically. We find that the couplings indeed flow as described for deviations below and above the attractive plane.

\section{Correlation length and gap scaling}\label{sec:corr}

\begin{figure}
\includegraphics[width=\columnwidth]{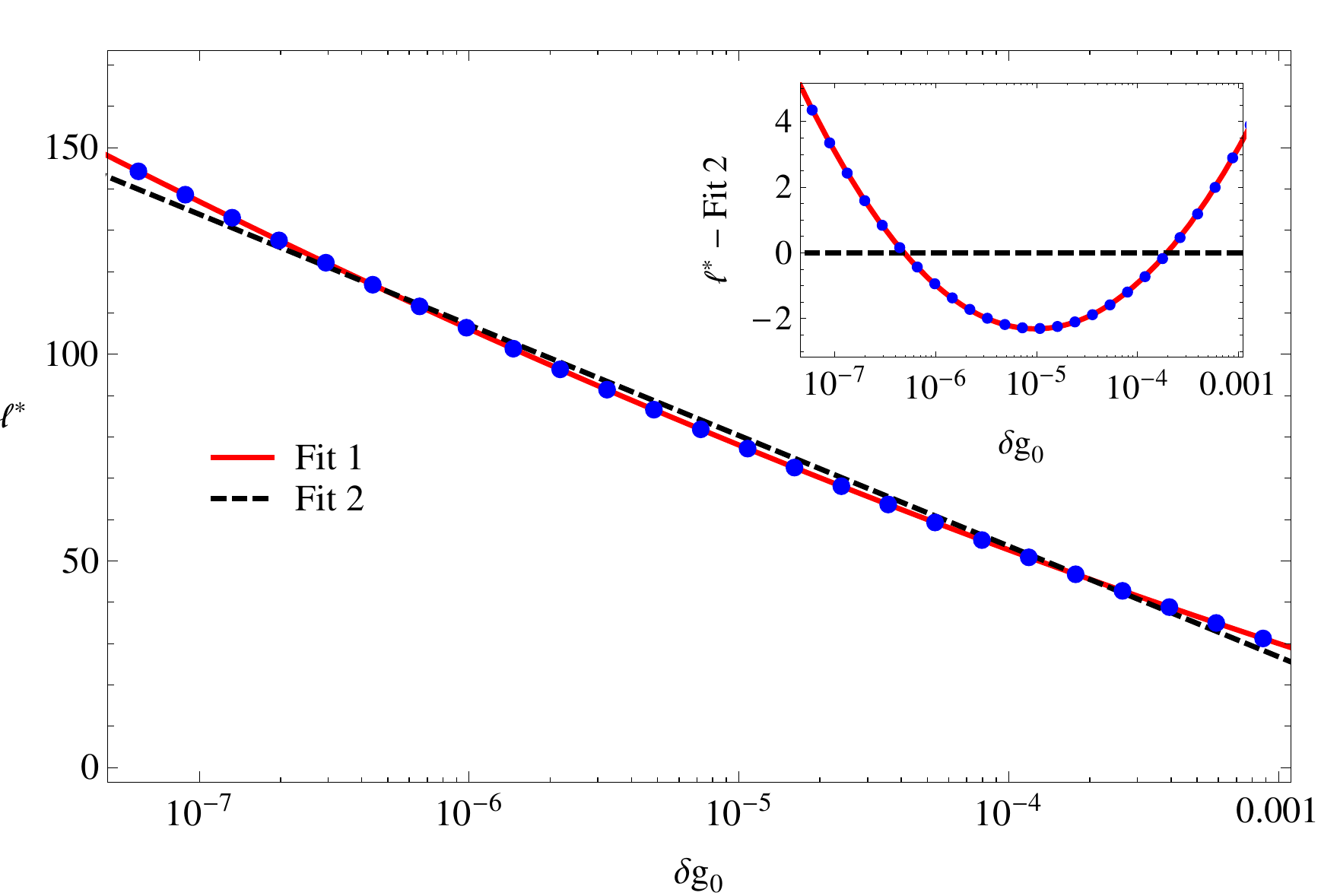}
\caption{We show the results of the numerical computation of the renormalization scale $\ell^*$ as a function of the deviation, $\delta g_0$, from the special line in the $g$-direction (blue dots). The red curve (fit 1) is a fit to the function $a \ln(\delta g_0)^2+b \ln(\delta g_0) +c$, with fit parameters $a\approx 0.26, b\approx5.70, c\approx -21.62$. For comparison the black, dashed curve (fit 2) is a fit to the function $b \ln(\delta g_0) +c$, with fit parameters $b\approx 11.62, c\approx -53.50$. In the inset we show the difference of the data and fit1 with fit 2. Clearly, the red curve is an excellent fit, while the black curve does not describe the scaling of $\ell^*$ that we observe. \label{fig:rgscale}}
\end{figure}

The intricate flow diagram around the fixed point and, in particular, the slow flow towards the fixed point have consequences on the physics of the multicritical point. In this section we compute how the correlation length diverges as the multicritical point is approached from the charge gapped phase, i.e. the correlation length associated to the $U(1)$ sector. Let us define $\xi_0$ as the value of the correlation in the gapped phase, that is far from the multicritical point, where the couplings are order one. It follows that 
\bea
\xi( g_i ) = \xi_0 e^{\ell^*}, \nonumber
\eea 
where $\ell^*$ is defined by $g_i(\ell^*)=g_i^*=\mathcal{O}(1)$, i.e. the renormalization scale at which the couplings become order one. To compute $\ell^*$ we start at a point on the special line and add a small perturbation, $\delta g_0$, that takes us above the attractive plane. This perturbation now grows under the RG flow according to (\ref{flowpert}). We now define the renormalization scale $\ell_1$ as the scale at which this perturbation reaches the size where the regime of validity of the linearized flow equations around the special line breaks down. More precisely, we define $\ell_1$ by
\bea
\delta g(\ell_1)=\lambda(\ell_1) \Rightarrow & & \nonumber\\
\delta g_0 e^{\frac{2}{\lambda_0}\sqrt{\lambda_0^2 \ell_1 + v_F^2}-\frac{2v_F}{\lambda_0}} &=& \frac{\lambda_0 v_F}{\sqrt{\lambda_0^2 \ell_1 + v_F^2}}. \nonumber
\eea
Solving this equation for $\ell_1$ gives
\bea
\ell_1 = \frac{1}{4} \left( \ln \left[ \frac{\lambda_0}{\delta g_0} \right] \right)^2 + \mathcal{O} \left( \ln \left[\frac{\lambda_0}{\delta g_0} \right] \right) . \nonumber
\eea
At this point, all the couplings are of the same order and validity of the linearization of the flow equation around the special line breaks down. However, since $g$ is increasing and $\lambda$ is decreasing, the flow will continue to a regime where the flow equations of the decoupled fixed point are valid, namely when $\lambda$ is sufficiently small compared to $g$. We will ignore this part of the flow trajectory for now -this will be justified below- and continue the trajectory as governed by the decoupled fixed point. The flow in $g$ then obeys
\bea
g(\ell)=\frac{g_0}{1-2 \pi \ell g_0/v_F}.
\eea
Defining $\ell_2$ as the RG scale at which $g$ is order one starting from $g(\ell_1)$, we obtain
\bea
& &1\sim g(\ell_2) = \frac{g(\ell_1)}{1-2 \pi \ell_2 g(\ell_1)/v_F} \nonumber\\
&\Rightarrow& \ell_2 \approx \frac{1}{2 \pi g(\ell_1)/v_F}\sim \sqrt{\ell_1} \nonumber
\eea
We see that $\ell_2 \ll \ell_1$, this difference is due to the fact that the flow in $g$ is now much faster than close to the special line. It is now clear that the part of the flow trajectory that we ignored is also negligible compared to the first part of the trajectory. We thus conclude that $\ell^* \approx \ell_1$ and, consequently,
\bea
\xi(\delta g_0) = \xi_0 e^{\frac{1}{4} \left( \ln \left[ \frac{\lambda_0}{\delta g_0} \right] \right)^2}.
\eea 
We verified this scaling of the correlation length by also solving the full flow equations numerically. We obtain a good fit of $\ell^*$ with the function $a \ln(\delta g_0)^2+b \ln(\delta g_0) +c$, where $a, b, c$ are fit parameters (see figure \ref{fig:rgscale}). Moreover, we extract $a \approx 0.26$, which is in good agreement with the expected value of $1/4$. Note that this coefficient of the leading term in $\ell^*$ is universal, i.e. independent of the details of how one approaches the multicritical point.


\section{Useful integrals}\label{app:int}
The integrals that we encounter when computing one- and two-loop diagrams are:
\bea
I_{1}(w, \bar{w})&= & \int\frac{d\omega dk}{(2\pi)^{2}}\frac{1}{\bar{z}}\frac{1}{z\pm w}\\
&=&\frac{1}{4\pi v}\Big[\frac{2}{\epsilon}-\log(|w|^{2})+\dots \Big],\nn
I_{2}(w, \bar{w})&= & \int\frac{d\omega dk}{(2\pi)^{2}}\frac{1}{\bar{z}}\frac{\bar{z}\pm\bar{w}}{z\pm w}\\
&= &\pm\bar{w}\frac{1}{4\pi v}\Big[\frac{2}{\epsilon}-\log(|w|^{2})+\dots\Big],\nn
I_{3}(w, \bar{w})&= & \int\frac{d\omega dk}{(2\pi)^{2}}\frac{1}{z}\frac{1}{z\pm w}=-\frac{1}{4\pi v}\Big(\frac{\bar{w}}{w}+1\Big), \label{I3}
\eea
where $z=\imath\omega+vk$ and $w=\imath\Omega+vq$. Furthermore, $\varepsilon$ is the dimensional regularization, $d=2-\varepsilon$, and the dots denote finite terms as $\varepsilon, |w|\to 0$. We point out that the last integral is not really well defined, but can be fixed on physical grounds. We first explain why the integral is ill-defined. Using Feynman parametrization and a change of variables, $r \exp(\imath \phi)=z+u w$, where $u$ is the Feynman parameter, the integral can be brought into the following form:
\bea
\int_0^1 d u \int \frac{dr d \phi}{(2\pi)^2 v} \frac{(r e^{-\imath \phi}-u \bar{w})(r e^{-\imath \phi}+(1-u) \bar{w})}{(r^2+(u-u^2)|w|^2)^2}.\nonumber
\eea
Carrying out the integral over the $\phi$-independent part of the integrand is now straight forward and gives the term proportional to $\bar{w}/w$ in (\ref{I3}). Naively, one might think that the $\phi$-dependent part of the integrand does not contribute, since the integral over $\phi$ gives zero. However, if one would carry out the $r$-integral first, the result is divergent. This subtlety can be resolved as follows. We note that this integral appears when one computes the density-density correlator. Using the fact that this correlator is known explicitly from bosonization we can fix the ill-defined integral. Let us provide some details. We have
\bea
\partial_x \Phi^{\pm} =\sqrt{2} \pi \sum_{a=1,2} \rho_{a \pm},   \nonumber
\eea
where the $\pm$ refers to left and right moving modes, respectively, and $\rho_{a \pm}=\psi_{a \pm}^{\dagger}\psi_{a \pm}$. The fermion density-density correlator is thus directly related to
\bea
\langle q \Phi^{\pm}_{\Omega,q} q' \Phi^{\pm}_{\Omega',q'} \rangle &=& \delta_{q+q'}\delta_{\Omega+\Omega'} \frac{-2\pi q}{\pm \imath \Omega+v q} .\label{Boscorr}
\eea
Computing this correlator for the left movers in the fermionic language gives
\bea
\langle q \Phi^-_{\Omega,q} q' \Phi^-_{\Omega',q'} \rangle = 2\pi^2 \sum_{a} \langle \rho_{a-} (\Omega,q) \rho_{a-} (\Omega',q') \rangle \nn
= \delta_{q+q'}\delta_{\Omega+\Omega'} 2\pi^2 \sum_a \int   \frac{d\omega dk}{(2\pi)^2} \Big( \frac{1}{z (z+w)} \Big), \nonumber
\eea
with $z,w$ as defined above. Since the ill-defined integral appears in the physically observable fermion density-density correlator, we can exclude the possibility that the integral is infinite. So let us assume $I_3(w, \bar{w})=-1/(4\pi v)(\bar{w}/w+f(w,\bar{w}))$, with $f$ some function. We then find
\bea
\langle q \Phi^-_{\Omega,q} q' \Phi^-_{\Omega',q'} \rangle =- \delta_{q+q'}\delta_{\Omega+\Omega'} \frac{\pi}{v}\Big(\frac{- \imath \Omega+vq}{\imath \Omega+vq}  + f(w,\bar{w}) \Big). \nonumber
\eea
Equating this with (\ref{Boscorr}) we find $f(w,\bar{w})=1$.

\end{document}